\documentclass[11pt,a4paper]{article}

\pdfoutput=1
\usepackage{jheppub}
\usepackage{simplewick}
\usepackage{thumbpdf}
\usepackage{amsmath}
\usepackage{amssymb}
\usepackage{array}
\usepackage{verbatim}
\usepackage{cancel}
\usepackage[hang,small]{caption2}
\usepackage{subfigure}
\newcommand\mycom[2]{\genfrac{}{}{0pt}{}{#1}{#2}}

\newcommand{\be}{\begin{equation}}
\newcommand{\ee}{\end{equation}}

\title{Bremsstrahlung function, leading L\"{u}scher correction at weak coupling and localization}
\author{Marisa Bonini$^{\bf a}$,}
\author{Luca Griguolo$^{\bf a}$,}  
\author{Michelangelo Preti$^{\bf a}$}
\author{and Domenico Seminara$^{\bf b}$}
\affiliation{$^{\bf a}$ Dipartimento di Fisica e Scienze della Terra, Universit\`a di Parma and INFN Gruppo Collegato di Parma, Viale G.P. Usberti 7/A, 43100 Parma, Italy}
\affiliation{$^{\bf b}$ Dipartimento di Fisica, Universit\`a di Firenze and INFN Sezione di Firenze, via G. Sansone 1, 50019 Sesto Fiorentino, Italy}
\emailAdd{marisa.bonini@fis.unipr.it} 
\emailAdd{luca.griguolo@fis.unipr.it} 
\emailAdd{michelangelo.preti@fis.unipr.it} 
\emailAdd{domenico.seminara@fi.infn.it} 

\abstract{We discuss the near BPS expansion of the generalized cusp anomalous dimension with 
L units of R-charge. Integrability provides an exact solution, obtained by solving a general TBA 
equation in the appropriate limit: we propose here an alternative method based on 
supersymmetric localization. The basic idea is to relate the computation to the vacuum 
expectation value of certain 1/8 BPS Wilson loops with local operator insertions along the 
contour. These observables localize on a two-dimensional gauge theory on $S^2$, opening the 
possibility of exact calculations. As a test of our proposal, we reproduce the leading L\"{u}scher 
correction at weak coupling to the generalized cusp anomalous dimension. This result is also 
checked against a genuine Feynman diagram approach in ${\cal N}=4$ Super Yang-Mills theory.} 

\preprint{}


\begin{document}
\maketitle

\section{Introduction}
In the last few years we have observed  surprising applications of powerful quantum field theoretical 
techniques to ${\cal N}=4$ Super Yang-Mills (SYM) theory. In particular the investigation of integrable structures \cite{Beisert:2010jr} and the use of supersymmetric localization \cite{Pestun:2007rz} have produced a huge number of results that are beyond perturbation theory and can be successfully compared with AdS/CFT expectations. 
For example the exact spectrum of single trace operator scaling dimensions can be obtained through TBA/Y-system equations   \cite{Arutyunov:2009zu,Arutyunov:2009ur,Gromov:2009tv,Bombardelli:2009ns} or, more efficiently, using the approach of the Quantum Spectral Curve (QSC) \cite{Gromov:2013pga}. Supersymmetric Wilson loops and classes of correlation functions can be instead computed exactly via supersymmetric localization, that exploit the BPS nature of the related observables \cite{Erickson:2000af,Drukker:2000rr,Drukker:2006ga,Drukker:2007dw,Drukker:2007qr,Pestun:2009nn,Giombi:2012ep,Giombi:2009ds}. It is certainly interesting to understand the relation between these two approaches, when the same quantities can be computed in both ways. 
\noindent

An important object, appearing in gauge theories is the cusp 
anomalous dimension $\Gamma(\varphi)$, that was originally introduced in \cite{Polyakov:1980ca,Korchemsky:1987wg} as the ultraviolet
divergence of a Wilson loop with Euclidean cusp angle $\varphi$.  In supersymmetric theories $\Gamma(\varphi)$ is not BPS, due to the presence of the cusp, and defines in the light-like limit $\varphi\to i\infty$ an 
 universal observable: exact results have been derived in ${\cal N}=4$ SYM through integrability
 \cite{Beisert:2006ez} that match both weak coupling expansions \cite{Bern:2006ew} and string computations describing 
 the strong coupling behavior \cite{Gubser:2002tv}. A related TBA approach, that goes beyond the light-like limit, was later proposed \cite{Correa:2012hh,Drukker:2012de}: the cusp anomalous dimension can be generalized 
 including an $R$-symmetry angle $\theta$ that controls the coupling of the scalars to the 
 two halves of the cusp \cite{Drukker:2011za}. The system interpolates between BPS configurations 
 (describing supersymmetric Wilson loops) and generalized quark-antiquark potential:
 exact equations can be written applying integrability, and have been checked successfully at three loops \cite{Correa:2012hh}. 
 For a recent approach using the QSC see \cite{Gromov:2015dfa}. Moreover one can use localization in a suitable limit to obtain the exact form of the infamous Bremsstrahlung function \cite{Correa:2012at}, that controls the near-BPS behavior of the cusp anomalous dimension (see also \cite{Fiol:2012sg} for a different derivation). The same result has been later directly recovered from the TBA  equations \cite{Gromov:2012eu,Gromov:2013qga} and QSC method \cite{Sizov:2013joa,Gromov:2015dfa}. It is clear that the generalized cusp anomalous dimension $\Gamma(\theta,\varphi)$ represents, in ${\cal N}=4$ SYM, a favorable playground in which the relative domains of techniques as integrability and localization overlap.

\noindent
The fundamental step that allowed to derive exact equations for $\Gamma(\theta,\varphi)$ from 
integrability was to consider the cusped Wilson loop with the insertion of $L$ scalars 
$Z = \Phi_1 + i\Phi_2$ on the tip. Importantly the scalars appearing in $Z$ should be 
orthogonal to the combinations that couple to the Wilson lines forming the cusp. The 
anomalous dimension  $\Gamma_{L}(\theta,\varphi)$ of the corresponding Wilson loop 
depends on $L$-unit of R-charge and a set of exact TBA equations can be written by 
for any value of $L$, $\theta$ and $\varphi$: setting $L=0$ one of course recovers 
$\Gamma(\theta,\varphi)$. When $\varphi=\pm\theta$ the operator becomes BPS and its near-BPS
 expansion in $(\varphi-\theta)$ can be studied directly from the integrability equations. The leading coefficient in this expansion is called the  
 Bremsstrahlung function when $L = 0$ and its expression has been derived to all loops from localization \cite{Correa:2012at}. For $\theta= 0$ it was reproduced at any coupling in \cite{Gromov:2012eu} by the analytic solution of the TBA, which also produced a new prediction for arbitrary $L$. 
 The extension to the case with arbitrary $(\theta-\varphi)$ and any $L$ is presented in \cite{Gromov:2013qga}. The near-BPS results for $L\geq1$ can be rewritten as a matrix model partition function whose classical limit was investigated in \cite{Sizov:2013joa} giving the corresponding classical spectral curve. A related approach based on a QSC has been proposed in \cite{Gromov:2015dfa}, reproducing the above results. 

In this paper we generalize the localization techniques of \cite{Correa:2012at} to the case of 
$\Gamma_{L}(\theta,\varphi)$, checking the prediction from integrability. 
Having an alternative derivation of the results may shed some light on the relations between 
these two very different methods. We consider here the same 1/4 wedge BPS Wilson loop used 
in \cite{Correa:2012at}, with contour lying on an $S^2$ subspace of $\mathbb{R}^4$ or $S^4$, 
inserting at the north and south poles, where the BPS cusp ends, $L$ "untraced" chiral 
primaries of the type introduced in \cite{Drukker:2009sf,Giombi:2009ds}. The analysis of the 
supersymmetries presented in \cite{Pestun:2009nn} 
shows\footnote{Actually in \cite{Giombi:2009ds} it was proved that a general family 
of  supersymmetric Wilson loops on $S^2$ and the local operators
 $O^L(x)={\rm tr}(\Phi_n+i\Phi_B)^L$ share two preserved supercharges.} that the system is 
 BPS. Following exactly the same strategy of \cite{Correa:2012at} we can relate the near 
 BPS limit of $\Gamma_{L}(\theta,\varphi)$ to a derivative of the wedge Wilson loop with local 
 insertions and define a generalized Bremsstrahlung function $\mathcal{B}_L(\lambda,\varphi)$ as in 
 \cite{Gromov:2012eu}. Our central observation is that the combined system of our Wilson 
 loop and the appropriate local operator insertions still preserves the relevant 
 supercharges to apply the localization procedure of \cite{Pestun:2009nn} 
and the computation of $\mathcal{B}_L(\lambda,\varphi)$ should be therefore possible in this framework. As usual supersymmetric  localization reduces the computation of BPS observables on a lower dimensional theory (sometimes a zero-dimensional theory $i.e$ a matrix model) where the fixed points of the relevant BRST action live. For general correlation functions of certain 1/8 BPS Wilson loops and
local operators inserted on a $S^2$ in space-time, localization reduces ${\cal N}=4$ SYM to a 2d
Hitchin/Higgs-Yang-Mills theory, that turns out to be equivalent to the two-dimensional pure Yang-Mills theory (YM$_2$) on $S^2$ in its zero-instanton sector\footnote{Truly to say, to complete the proof of \cite{Pestun:2009nn} one should still evaluate the one-loop determinant for the field fluctuations in the directions normal to the localization locus. However, there are many reasons to believe that such determinant is trivial.}. 
In particular the four-dimensional correlation functions are captured by a perturbative calculation in YM$_2$ on $S^2$. In two dimensions, the preferred gauge choice is the light-cone gauge, since then there are no interactions: the actual computations are drastically simplified thanks to the quadratic nature of the effective two-dimensional theory. This sort of dimensional reduction has passed many non-trivial checks \cite{Giombi:2009ms,Bassetto:2008yf,Young:2008ed,Bassetto:2009rt,Bassetto:2009ms,Bonini:2014vta}, comparing the YM$_2$ results both at weak coupling (via perturbation theory in ${\cal N}=4$ SYM) and strong coupling (using AdS/CFT correspondence). 

\noindent
Here we investigate the near-BPS limit of $\Gamma_{L}(\theta,\varphi)$ using perturbative YM$ _2$ on the sphere: more precisely we attempt the computation of $\mathcal{B}_L(\lambda,\varphi)$. The problem of resumming in this case the perturbative  series in YM$_2$ is still formidable (we will come back on this point in the conclusions) and we limit ourselves here to the first non-trivial perturbative order. In so doing we recover, in closed form, the leading L\"{u}scher correction to the ground state energy of the open spin chain, describing the system in the integrability approach ($\lambda=g^2N$)
\begin{equation}
\Gamma_L(\theta,\varphi)\simeq 
(\varphi-\theta) 
\frac{(-1)^{L}\lambda^{L+1}}{4\pi(2L+1)!}B_{2L+1}\left(\frac{\pi-\varphi}{2\pi}\right)\,,
\end{equation}
where $B_{n}(x)$ are the Bernoulli polynomials.
This correction was computed in \cite{Correa:2012hh, Drukker:2012de}, following directly from the TBA equations,  and checked from the general formulae in \cite{Gromov:2012eu,Gromov:2013qga}. Interestingly it was also reproduced in \cite{Correa:2012nk} by a direct perturbative calculation in some apparently unrelated amplitude context. We think that our result not only represents a strong check that our system computes the generalized Bremsstrahlung function but also suggests some relations between two-dimensional diagrams and four-dimensional amplitudes.

The plan of the paper is the following: in Section 2 we present the construction of the wedge Wilson loop in ${\cal N}=4$ SYM with the relevant operator insertions. 
We discuss the BPS nature of these observables and their mapping to two-dimensional Yang-Mills theory on $S^2$ in the zero-instanton sector. 
The relation with the generalized Bremsstrahlung function is also established. 
In Section 3 we set up the perturbative calculation of the first non-trivial contribution to the near-BPS limit of $\Gamma_{L}(\theta,\varphi)$, first considering the case $L=1$ and then presenting the computation for general $L$. 
We recover the weak coupling contribution to the L\"{u}scher term as expected. 
In Section 4 we perform the $L=1$ computation using Feynman diagrams in ${\cal N}=4$ SYM: this is a check of our construction and also exemplifies the extreme complexity of the conventional perturbative calculations. 
In Section 5 we present our conclusions and perspectives, in particular discussing how the full result for  $\mathcal{B}_L(\lambda,\varphi)$ could be recovered  and its matrix model formulation should arise from YM$_2$ on the sphere. 
Appendix A is instead devoted to some technical aspects of YM$_2$ computation.

\section{Loop operators in  ${\cal N}=4$ SYM and  YM$_2$: an alternative route to  the generalized  Bremsstrahlung function}
For the $\mathcal{N}=4$ SYM theory an  interesting and infinite  family of loop  operators  which share, independently of the contour,  $\frac{1}{8}$ of the original supersymmetries of the theory were constructed in \cite{Drukker:2007dw,Drukker:2007qr}.  
They compute the holonomy of a generalized connection
combining  the gauge field $(A_\mu)$, the scalars $(\Phi^I)$ and the invariant forms $(\epsilon_{\mu\nu\rho} dx^\nu x^\rho)$ on $S^2$ as follows 
(we follow the conventions of \cite{Giombi:2012ep})
\begin{equation}\label{WN4}
\mathcal{W}=\frac 1 N \text{Tr}\left[\mathrm{P}\!\exp{\left(\oint d\tau\left(\dot{x}^\mu A_\mu+i\epsilon_{\mu\nu\rho}
\dot{x}^{\mu}x^{\nu}\Phi^\rho\right)\right)}\right].
\end{equation}
Above we have parameterized the sphere  in terms of three spatial cartesian coordinates obeying the constraint $x_1^2+x_2^2+x_3^2=1$. The normalization $N$ is the rank of the unitary gauge group $U(N)$. 

In \cite{Drukker:2007dw,Drukker:2007qr}   it was also  argued that the expectation value of this class of observables was captured by the matrix model governing the zero-instanton sector of two dimensional Yang-Mills on the sphere:
\be
\label{matrixmodel}
\begin{split}
\langle\mathcal{W}^{(2d)}\rangle=&\frac{1}{\mathcal{Z}}\int DM \frac{1}{N} \mathrm{Tr}\left(e^{i M}\right)\exp\left(-\frac{A}{g^2_{2d} A_1 A_2}\mathrm{Tr}(M^2)\right)=\\
=&\frac{1}{N}  L^1_{N-1} \left(  \frac{g_{2d}^2 A_1 A_2}{2A}\right)\exp
\left(- \frac{g_{2d}^2 A_1 A_2}{4A}\right),
\end{split}
\ee
where $A_1$, $A_2$ are the areas singled out by the Wilson loop and $A= A_1 + A_2=4\pi$ is the total area of the sphere. 
The result in $\mathcal{N}=4$ SYM is then obtained with the following map
\be
\label{coup2D4D}
g_{2 d}^2 \mapsto - \frac{2g^2}{ A}.
\ee
Above $g_{2d}$ and $g$ are respectively  the two  and four dimensional  Yang-Mills coupling constants. 

This intriguing connection between YM$_2$ and $\mathcal{N}=4$ SYM was put on solid ground  by  Pestun in \cite{Pestun:2007rz,Pestun:2009nn}. By means of localization  techniques, he was able to argue that the four-dimensional  $\mathcal{N }= 4$ 
Super Yang-Mills theory on $S^4$ for this class of observables reduces to the two-dimensional constrained Hitchin/Higgs-Yang-Mills (cHYM) theory on $S^2$. The Wilson loops in the cHYM theory  are  then shown to be given by those of YM$_2$ in the zero instanton sector. These results were extended to the case of correlators of  Wilson loops  in \cite{Giombi:2009ds,Bassetto:2008yf} and to the case of correlators of Wilson loop with chiral primary operators in \cite{Giombi:2009ds,Giombi:2012ep}.

The knowledge of the exact expectation value  for this family of  Wilson loops has been a powerful tool to test  the AdS/CFT correspondence in different regimes. In particular,  here, we want to investigate the connection, originally discussed in  \cite{Correa:2012at}, between these  observables and the so-called Bremsstrahlung function.

Consider,  in fact, the generalized cusp $\Gamma(\theta,\varphi)$  defined in  \cite{Drukker:2011za}, namely  the coefficient of the logarithmic divergence for a Wilson line that makes a turn by an angle $\varphi$ in actual space-time and  by an 
angle $\theta$ in   the $R-$symmetry space of the theory. When $\theta =\pm \varphi$  the  Wilson line becomes  supersymmetric and  
$\Gamma(\theta,\varphi)$ identically  vanishes. In this limit  the operator is, in fact, a particular case of the $\frac{1}{4}$ BPS loops discussed by Zarembo in \cite{Zarembo:2002an}.  In \cite{Correa:2012at}  a simple and compact  expression for the first order deviation away from the BPS value was derived.  When $|\theta-\varphi|\ll 1$, we can write
\be
\label{Gammacusp}
\Gamma(\theta,\varphi)\simeq 
-(\varphi-\theta) \mathcal{H}({\lambda},\varphi)+O((\varphi-\theta)^2)
\ee
where
\be
\label{Hfunction}
\mathcal{H}({\lambda},\varphi)=\frac{2\varphi}{1-\frac{\varphi^2}{\pi^2}} \mathcal{B}(\tilde\lambda)\qquad 
\text{with}\quad \tilde\lambda=\lambda\left(1-\frac{\varphi^2}{\pi^2}\right).
\ee
In \eqref{Hfunction} $\mathcal{B}(\tilde{\lambda})$  is given by the Bremsstrahlung function of $\mathcal{N}=4$ SYM theory. 

The  expansion \eqref{Gammacusp}
 was obtained in \cite{Correa:2012at} by considering a small deformation of  the so-called   $\frac{1}{4}$ BPS wedge.  It is 
 a loop in the class \eqref{WN4} which consists of two meridians separated by an angle $\pi-\varphi$.    The analysis in   \cite{Correa:2012at} in particular shows that  $\mathcal{H}(\lambda,\varphi)$ can be computed as the logarithmic derivative of the expectation value
 of  the BPS wedge with respect to the angle $\varphi$
\begin{equation}\label{defBH}
\mathcal{H}(\lambda,\varphi)=-\frac{1}{2}
\partial_\varphi\log {\langle \mathcal{W}_{\rm wedge}(\varphi)\rangle}
=-\frac{1}{2}\frac{\partial_\varphi\langle \mathcal{W}_{\rm wedge}(\varphi)\rangle}{\langle \mathcal{W}_{\rm wedge}(\varphi)\rangle}.
\end{equation}
The quantity   $\langle \mathcal{W}_{\rm wedge}(\varphi)\rangle$ is given by the  matrix model \eqref{matrixmodel} with the replacement \eqref{coup2D4D}. In other words   $\langle \mathcal{W}_{\rm wedge}(\varphi)\rangle=
\langle \mathcal{W}_{\rm circle}(\tilde \lambda)\rangle$ where $\tilde\lambda$ is defined in \eqref{Hfunction}.

The results \eqref{Hfunction} and \eqref{defBH} were also recovered in   \cite{Gromov:2012eu,Gromov:2013qga} by solving the TBA  equations, obtained in \cite{Correa:2012hh},  for the cusp anomalous dimension  in the BPS limit. This second  approach
based on integrability  naturally led to consider a  generalization  $\Gamma_L(\theta,\varphi)$  for the cusp anomalous dimension, where one has inserted  the scalar operator $Z^L$  on the tip of  the cusp\footnote{ The scalar $Z$ is the holomorphic combination of two scalars, which do not couple to the Wilson line.}. In \cite{Gromov:2013qga} it was shown that  the whole family  $\Gamma_L(\theta,\varphi)$ admits an expansion of the type \eqref{Gammacusp} when 
$|\theta-\varphi|\ll 1:$
\be
\label{GammaLcusp}
\Gamma_L(\theta,\varphi)\simeq 
-(\varphi-\theta) \mathcal{H}_L(\lambda,\varphi)+O((\varphi-\theta)^2)
\ee
with
\be
\label{HLfunction}
\mathcal{H}_L(\lambda,\varphi)=\frac{2\varphi}{1-\frac{\varphi^2}{\pi^2}} 
\mathcal{B}_L(\lambda,\varphi)\,.
\ee
The function $\mathcal{B}_L(\lambda,\varphi)$  is  a generalization  of the usual  Bremsstrahlung function and its value for any $L$ in the large  $N$ limit was derived in \cite{Gromov:2013qga}.

Below, we want to show that expansion  \eqref{GammaLcusp} and in particular  the function $\mathcal{B}_L(\lambda,\varphi)$  can be evaluated exploiting the relation between  the  Wilson loops \eqref{WN4} and  YM$_2$ as done in \cite{Correa:2012at} for the case $L=0$.

\subsection{BPS wedge on $S^2$  with scalar insertions in $\mathcal{N}=4$ SYM}
The first step is to construct a generalization of  the $\frac{1}{4}$ BPS wedge  in $\mathcal{N}=4$ SYM, whose vacuum expectation value is still determined by a suitable observable in YM$_2$. 
 \begin{figure}[ht]
\centering
	\includegraphics[width=5cm]{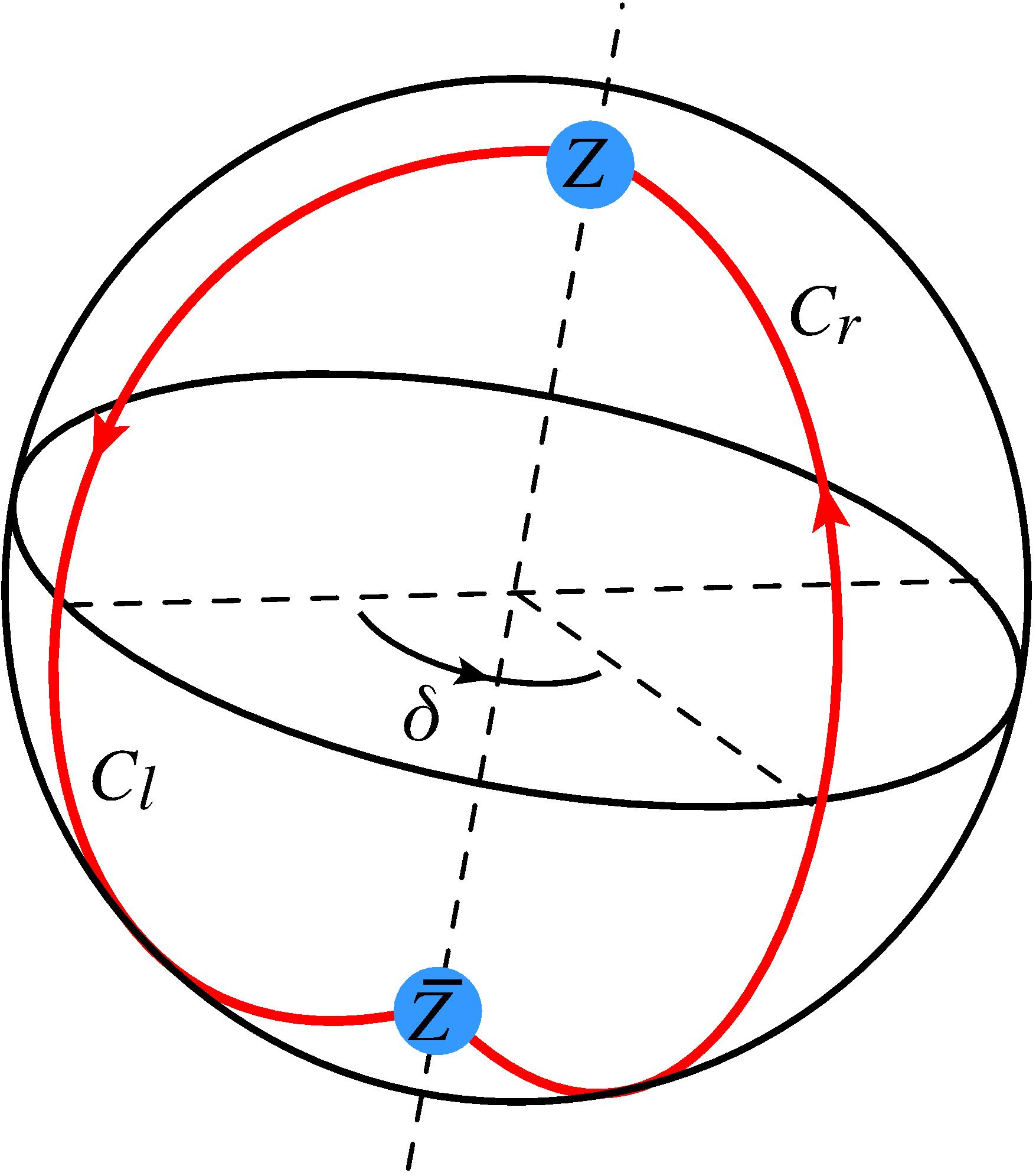}
	\caption{Pictorial description of the observable \eqref{ObsN4}:  the $\frac{1}{4}$ BPS wedge with  
	operator insertions
	in the north and the south pole.}
		\label{fig:spicchio1}
\end{figure}
We start by considering the  contour depicted in Figure \ref{fig:spicchio1}, namely  the wedge  composed by two meridians 
$C_l$ and $C_r$, the first in the $(x_1,x_3)$ plane and the second in a plane with
longitude angle $\delta$:
\begin{equation}\begin{split}
\label{xLxR}
&C_l\quad\mapsto\quad x_{l}^\mu=(\sin{\tau},0,\cos{\tau})\qquad\qquad\qquad\qquad\quad\;\; 0\leq\tau\leq\pi,\\
&C_r\quad\mapsto\quad x_{r}^\mu=(-\cos{\delta}\sin{\sigma},-\sin{\delta}\sin{\sigma},\cos{\sigma})\qquad \!\!\pi\leq\sigma\leq 2\pi.
\end{split}\end{equation}
On the two meridians the Wilson loop couples  to two different combinations of scalars.  
Indeed the effective gauge connections on the two sides are given by
\begin{equation}
\mathcal{A}_l= \dot{x}_l^\mu A_\mu-i
\Phi^2\,,\quad\quad\quad \mathcal{A}_r= \dot{x}_r^\mu A_\mu
-i\sin{\delta}\Phi^1+i\cos{\delta}\Phi^2.
\end{equation}
We now consider the insertion of local operators in the loop and focus our attention on those introduced in \cite{Drukker:2009sf} and studied in detail in \cite{Giombi:2012ep}. They are
given by:
\begin{equation}
\label{CPO}
\mathcal{O}_L(x)=\left( x_\mu \Phi^\mu+i \Phi^4\right)^L ,
\end{equation}
where $x_\mu$ are the cartesian coordinate of a point on $S^2$ and the index $\mu$ runs from $1$ to $3$. 
Any system of  these operators preserves at least four supercharges. When the Wilson loops \eqref{WN4} are also present, the combined system is generically invariant under two supercharges $\left(\frac{1}{16}~ \rm BPS\right)$ \cite{Giombi:2009ds}. Here we choose to insert  two  of these operators: one in the north pole [$x_{N}^\mu=(0,0,1)$]  and  the other in the south pole [$x_{S}^\mu=(0,0,-1)$]. In these special positions they reduce to the holomorphic and the anti-holomorphic combination
of two of the scalar fields which do not couple to the loop  
\begin{equation}
\label{Insertions}
 \mathcal{O}_L (x_N)=\left(\Phi^3+i \Phi^4\right)^L\equiv Z^L\,,\quad\quad\quad
 \mathcal{O}_L(x_S)=\left(-\Phi^3+i \Phi^4\right)^L\equiv (-1)^L\bar{Z}^L.
\end{equation}
Then the generalization of  the $\frac{1}{4}$ BPS wedge  which is supposed to capture  the generalized  Bremsstrahlung function
$\mathcal{B}_L(\lambda,\varphi)$ is simply given by
\begin{equation}
\label{ObsN4}
\mathcal{W}_{L}(\delta)=\text{Tr}\left[Z^L~\mathrm{P}\!\exp\left(\int_{C_l}\mathcal{A}_l d\tau\right) ~\bar Z^L\,\mathrm{P}\!\exp\left(\int_{C_r}\mathcal{A}_r 
d\sigma\right)\right].
\end{equation}
The operators \eqref{ObsN4} preserve $\frac{1}{8}$ of the  supercharges. Applying the same argument  given in  \cite{Correa:2012at}, one can argue (since the relevant deformation never involves the poles) that 
\begin{equation}\label{defBHL}
\mathcal{H}_L(\lambda,\varphi)=\frac{2\varphi}{1-\frac{\varphi^2}{\pi^2}}\,\mathcal{B}_L(\lambda,\varphi)=\left.\frac{1}{2}
\partial_\delta \log {\langle \mathcal{W}_{ L}(\delta)\rangle}\right|_{\delta=\pi-\varphi}.
\end{equation}

\subsection{The wedge on $S^2$  with field strength insertions in YM$_2$}
Next we shall construct a {\it putative} observable in  YM$_2$, which computes the vacuum expectation of $\mathcal{W}_L(\delta)$.
 \begin{figure}[ht]
\centering
	\includegraphics[width=9cm]{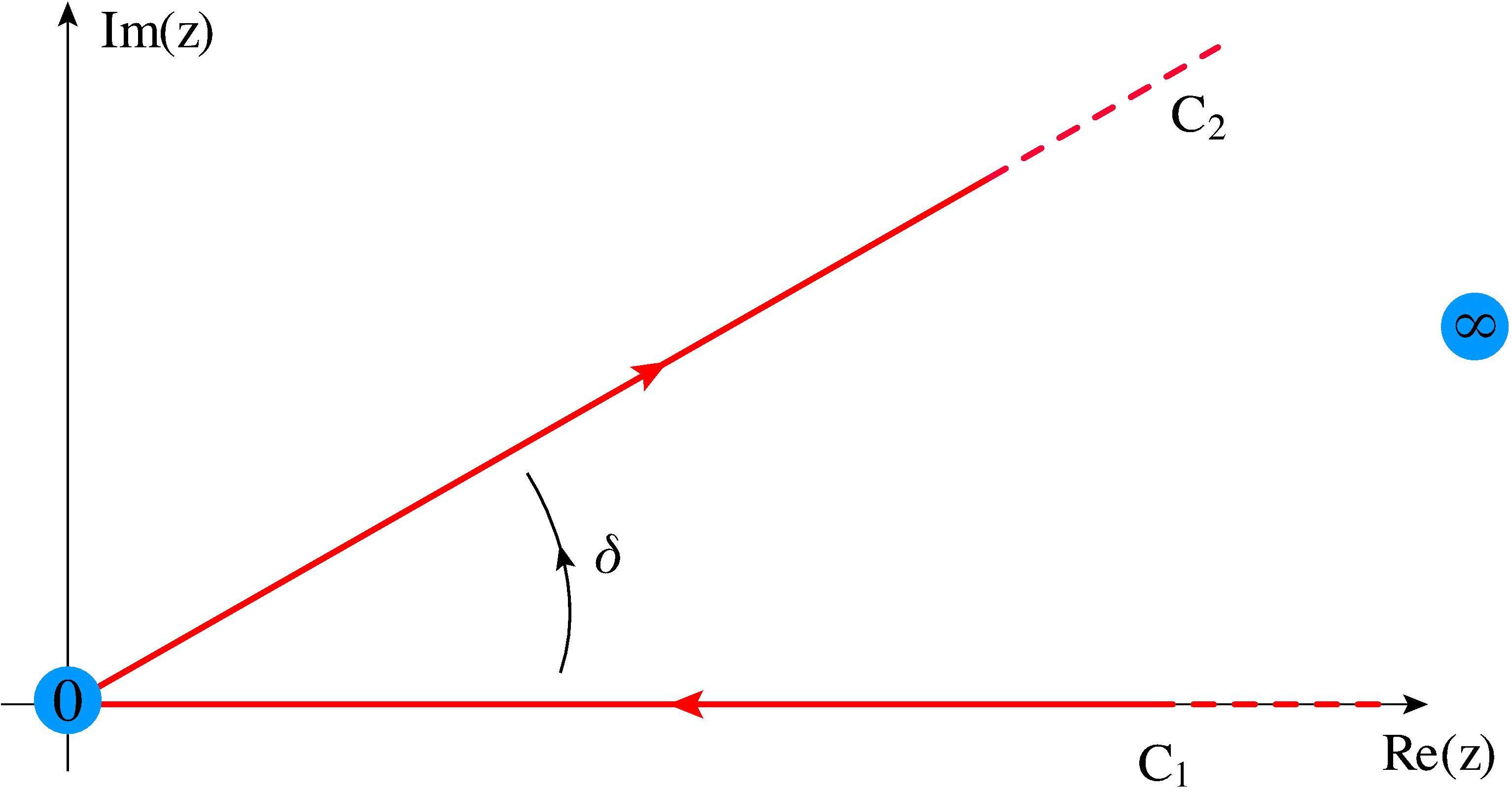}
	\caption{Wedge on $S^2$ in stereographic coordinates. The blue blobs denote the operator insertions.}
		\label{fig:spicchio2d}
\end{figure}
Following \cite{Drukker:2007qr}, we map the Wilson loop of $\mathcal{N}=4$ SYM into a loop operator of YM$_2$
defined along the same contour $C$  on $S^2$.  Since we use the complex stereographic  coordinates $z=x+i y$ to parametrize the sphere, 
the wedge will appear as an infinite cusp on the plane where the origin represents  the north pole, while  the infinity is identified with the south pole
 (see Figure \ref{fig:spicchio2d}).  The  two straight-lines $C_1$ and $C_2$  are then given by
 \begin{equation}\begin{split}
&C_1\quad\mapsto\quad 
z(t)=t\,\qquad\,\,\quad \;\;t\in [\infty,0]\,,\\
&C_2\quad\mapsto\quad 
w(s)=e^{i\delta}s\qquad\  \!\;s\in [0,\infty]\,.
\end{split}\end{equation} 
In these coordinates the  metric  on $S^2$  takes the usual conformally   flat form
\begin{equation}
ds^2=\frac{4dzd\bar{z}}{(1+z\bar{z})^2}\,.
\end{equation}
The wedge Wilson loop is mapped into
\begin{equation}
\label{W2wedge}
\mathcal{W}_C=\frac{1}{N} \text{Tr}\left[\mathrm{P}\!\exp\left(\oint_C d\tau\; 
\dot{z}\tilde{A}_z(z)\right)\right]\,,
\end{equation}
where $C=C_1\cup C_2$ (see Figure \ref{fig:spicchio2d}). In \eqref{W2wedge} we have used the notation   $\tilde{A}$,  to distinguish gauge field  in $d=2$ from its  counterpart $A$ in $d=4$. 

\noindent
The two dimensional companion of the operator  \eqref{CPO}  was  found in  \cite{Giombi:2012ep}  through a localization argument.
These local operators are mapped into powers of the  field strength $\tilde{F}$ of the two-dimensional gauge field $\tilde A$:
 \begin{equation}
\mathcal{O}_L(x)\mapsto \left(i~{}^*\!\tilde{F}(z)\right)^L\,.
\end{equation}
Here the $^*$ denotes the usual hodge dual on the sphere.  
Combining the above ingredients the  observable in YM$_2$ which computes the vacuum expectation value of \eqref{ObsN4} is
\begin{equation}
\label{ObsYM2}
{\mathcal{W}}_{L}^{(2d)}(\delta)=\text{Tr}\left[ \left(i~{}^*\!\tilde{F}(0)\right)^L~\mathrm{P}
\!\exp\left(\int_{C_2}\dot{w}\tilde{A}_w ds\right) ~\left(i~{}^*\!\tilde{F}(\infty)\right)^L\,\mathrm{P}
\!\exp\left(\int_{C_1}\dot{z}\tilde{A}_z dt\right)\right].
\end{equation}
Constructing a matrix model for the observable \eqref{ObsYM2} in the zero instanton sector is neither easy nor immediate.  In fact  the usual topological Feynman  rules for computing quantities in YM$_2$ are tailored  for the case of  correlators of  Wilson loops.  
The insertion of local operators along the contour was not considered  previously and  the rules for this case are still missing. Therefore,
in the next section, to test the relation between the  function $\mathcal{H}_L(\lambda,\varphi)$ and  the vacuum expectation value of  \eqref{ObsYM2}
implied by \eqref{defBHL}, {\it i.e.}
\begin{equation}\label{defBHL2}
{
\mathcal{H}_L(\lambda,\varphi)=\frac{2\varphi}{1-\frac{\varphi^2}{\pi^2}}\,\mathcal{B}_L(\lambda,\varphi)=\left.\frac{1}{2}
\partial_\delta \log {\langle \mathcal{W}^{(2d)}_{ L}(\delta)\rangle}\right|_{\mycom{\delta=\pi-\varphi }{g_{2d}^2=-{2g^2}/{A}}},}
\end{equation}
we shall resort to standard perturbative techniques. 

\section{Perturbative computation of the L\"{u}scher term from YM$_2$ on the sphere}
In this Section we compute the first non-trivial perturbative contribution to the generalized  Bremsstrahlung function $\mathcal{B}_L(\lambda,\varphi)$, 
in the planar limit, using perturbation theory in YM$_2$. It coincides with 
the weak coupling limit of the L\"{u}scher term, describing wrapping corrections in 
the integrability framework. We believe it is instructive to present first the computation 
for $L=1$, where few diagrams enter into the calculation and every step can be followed explicitly. Then we turn to general $L$, exploiting some more sophisticated techniques to account for the combinatorics.
\subsection{General setting for perturbative computations on $S^2$}
For perturbative calculations on the two dimensional sphere, it is convenient to use  the holomorphic gauge
$\tilde{A}_{\bar{z}}=0$.
In this gauge the interactions vanish  and the relevant propagators are:
\begin{equation}\begin{split}\label{prop2d}
\langle(\tilde{A}_z)^i_j(z)(\tilde{A}_z)^k_l(w)\rangle&=-\frac{g^2_{2d}}{2\pi}\delta^i_l\delta^k_j
\frac{1}{1+z\bar{z}}\frac{1}{1+w\bar{w}}\frac{\bar{z}-\bar{w}}{z-w}\,,\\
\langle (i ^*\tilde{F}^i_j(z)) (i ^*\tilde{F}^k_l(w))\rangle&=
-\delta^i_l\delta^k_j\left(\frac{g^2_{2d}}{8\pi}-\frac{ig^2_{2d}}{4}(1+z\bar{z})^2\delta^{(2)}(z-w)\right)\,,\\
\langle (i ^*\tilde{F}^i_j(z)) (\tilde{A}_z)^k_l(w)\rangle&=-\frac{g^2_{2d}}{4\pi}\delta^i_l\delta^k_j
\frac{1}{1+w\bar{w}}\frac{1+z\bar{w}}{z-w}\,.
\end{split}\end{equation}
When computing the above propagators for points lying on the edge $C_1$  ($z=t $) and on the edge $C_2$ ($w=e^{i\delta} s$),
they reduce to 
\begin{align}\label{prop2dreal}
&\langle(\tilde{A}_z)^i_j(z)(\tilde{A}_z)^k_l(w)\rangle=-\frac{g^2_{2d}}{2\pi}\delta^i_l\delta^k_j
\frac{1}{1+t^2}\frac{1}{1+s^2}\,\frac{t-e^{-i\delta}s}{t-e^{i\delta}s}\,,\nonumber\\
&\langle (i ^*\tilde{F}^i_j(0) )(\tilde{A}_z)^k_l(z)\rangle=\frac{g^2_{2d}}{4\pi}\delta^i_l\delta^k_j
\frac{1}{1+t^2}\frac 1 t\,,\qquad 
\langle (i ^*\tilde{F}^i_j(0)) (\tilde{A}_z)^k_l(w)\rangle=\frac{g^2_{2d}}{4\pi}\delta^i_l\delta^k_j
\frac{1}{1+s^2}\frac{e^{-i\delta}}{s}\,,\nonumber\\
&\langle (i ^*\tilde{F}^i_j(\infty) )(\tilde{A}_z)^k_l(z)\rangle=-\frac{g^2_{2d}}{4\pi}\delta^i_l\delta^k_j
\frac{t}{1+t^2}\,, \qquad\!\!
\langle  (i ^*\tilde{F}^i_j(\infty)) (\tilde{A}_z)^k_l(w)\rangle=-\frac{g^2_{2d}}{4\pi}\delta^i_l\delta^k_j
\frac{e^{-i\delta}s}{1+s^2}\,.
\end{align}
These Feynman rules can be checked by computing the first two non trivial orders for the standard wedge 
in YM$_2$, {\it i.e.} $L=0$.
One quickly finds:
\begin{equation}\begin{split}
\langle\mathcal{W}^{(2d)}_{\text{1-loop}}\rangle=&\frac{g_{2d}^2 N}{2\pi}\left(\frac{\pi^2}{4}-\frac 12 (2\pi-\delta)\delta\right)\,,\\
\langle\mathcal{W}^{(2d)}_{\text{2-loop}}\rangle=&-\frac{g_{2d}^4 N^2}{96\pi^2}(2\pi-\delta)^2\delta^2\,.
\end{split}\end{equation}
Using \eqref{defBHL2} one gets
\begin{equation}\begin{split}
\mathcal{H}^{(1)}(\lambda,\varphi)=\frac{\lambda\varphi}{8\pi^2},\qquad\qquad\mathcal{H}^{(2)}(\lambda,\varphi)=-\frac{\lambda^2\varphi(\pi^2-\varphi^2)}{192\pi^4},
\end{split}\end{equation}
which is exactly the result of \cite{Correa:2012at}.

\subsection{Operator insertions of length $L=1$}
\begin{figure}[!h]
 \centering
  \subfigure[]
   {\includegraphics[width=5cm]{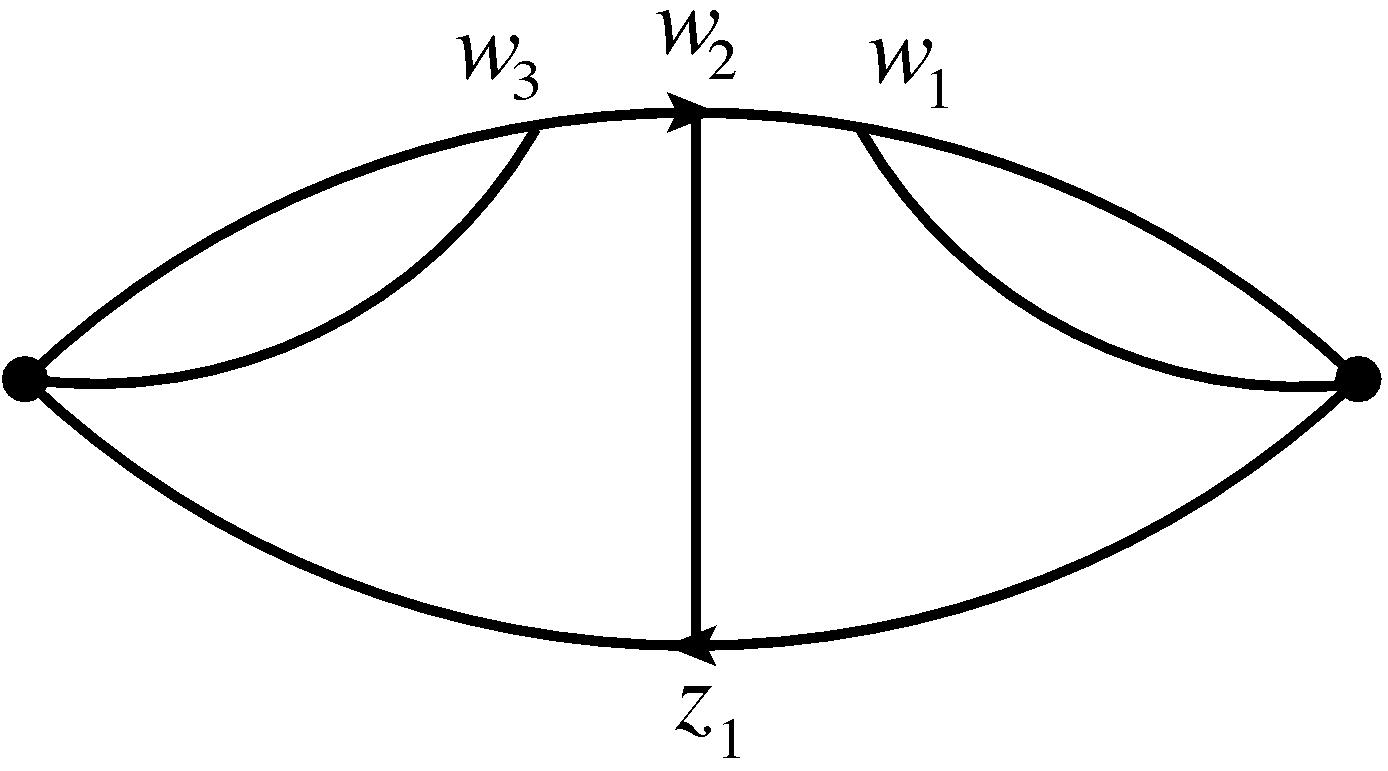}}
    \hspace{5mm}
    \subfigure[]
   {\includegraphics[width=5cm]{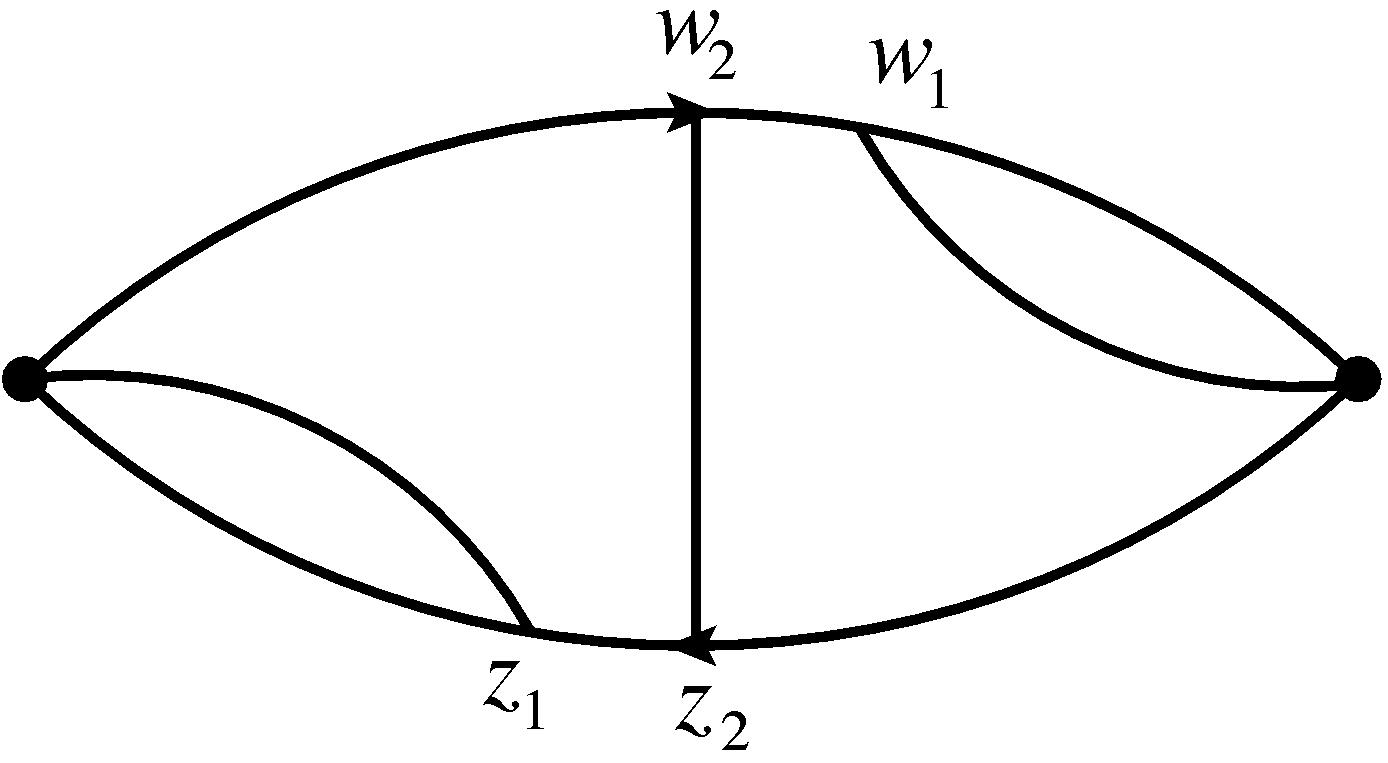}}
    \hspace{5mm}
 \subfigure[]
   {\includegraphics[width=5cm]{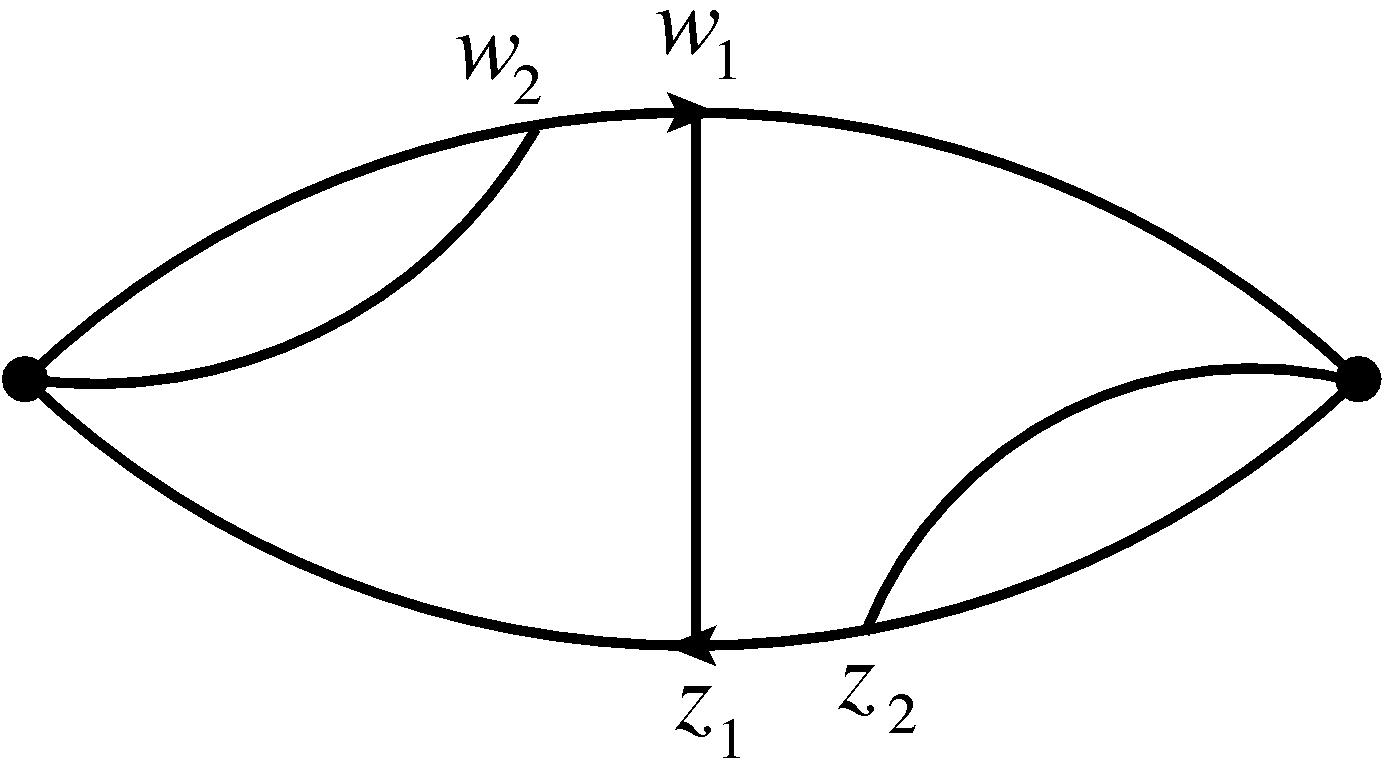}}
        \hspace{5mm}
 \subfigure[]
   {\includegraphics[width=5cm]{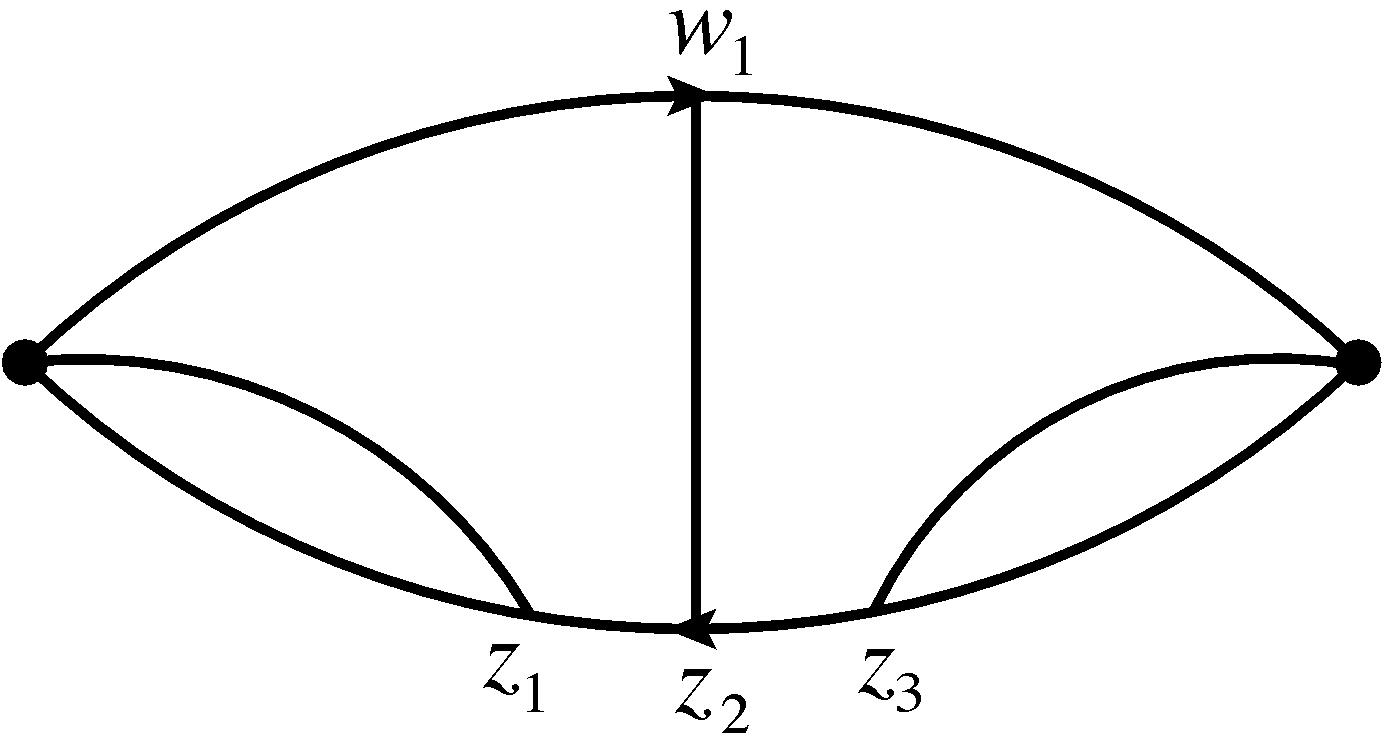}}
    \hspace{5mm}
 \caption{All the $\delta$-dependent diagrams at order $\lambda^2$ in $YM_2$}
  \label{fig:grafici2dd}
 \end{figure}
The first non-trivial contribution in this case appears at order $\lambda^2$: we need at least one propagator  connecting the two halves of the wedge, 
to carry the dependence on the opening angle $\delta$. Then we have to consider the effect of the operator insertions, which should be connected to the contour, respecting planarity, in all possible ways. The relevant choices are represented in Figure
\ref{fig:grafici2dd}.
We remark that every diagram has the same weight: in fact the operator we study 
is single-trace and every diagram arises from an unique set of Wick contractions.
Furthermore there are some obvious symmetries between the different diagrams, 
that imply the following representation:
\begin{equation}\begin{split}\label{intconF}
\text{(a)}+\text{(d)}&=2\frac{g_{2d}^6N^4}{32\pi^3}\int_0^\infty \!\!\!ds_1\!\!\int_0^{s_1}\!\!\!ds_3
\frac{1}{(s_1^2+1)(s_3^2+1)}\frac{s_1}{s_3}\,\mathcal{F}_\delta(\infty,0;s_3,s_1)\,,\\
\text{(b)}+\text{(c)}&=\frac{g_{2d}^6N^4}{32\pi^3}\int_\infty^0 \!\!\!dt_1\!\!\int_0^{\infty}\!\!\!ds_1
\frac{1}{(t_1^2+1)(s_1^2+1)}\frac{s_1}{t_1}\biggl[\mathcal{F}_\delta(\infty,t_1;0,s_1)+\mathcal{F}_{-\delta}(\infty,t_1;0,s_1)\biggl]\,,
\end{split}\end{equation}
where we have introduced the function $\mathcal{F}_\delta(a,b;c,d)$
\begin{equation}\label{defF}
\mathcal{F}_\delta(a,b;c,d)=\int_a^b dt\int_c^d ds \;\frac{1}{(t^2+1)(s^2+1)}\,\frac{e^{i\delta}t-s}{t-e^{i\delta}s}\,.
\end{equation} 
Using the identity $\frac{1}{(s^2+1)(t^2+1)}\frac{s}{t}=\frac14\partial_s\partial_t[\log{\left(s^2+1\right)}\log{\left(t^2/({t^2+1})\right)}]$ and integrating by parts, we obtain
\begin{align}\label{intparti}
\text{(a)}+\text{(d)}=&\frac{g_{2d}^6N^4}{32\pi^3}\Biggl[2\log^2\epsilon \mathcal{F}_\delta(\infty,0;0,\infty)
+\log\epsilon\int_0^\infty ds_1 
\log(s_1^2+1)\frac{d}{ds_1}\mathcal{F}_\delta(\infty,0;0,s_1)+\nonumber\\
&+\log\epsilon\int_0^\infty ds_3 
\log\left(\frac{s_3^2}{s_3^2+1}\right)\frac{d}{ds_3}\mathcal{F}_\delta(\infty,0;s_3,\infty)+\nonumber\\
&+\frac 12\int_0^\infty 
ds_1\log(s_1^2+1)\log\left(\frac{s_1^2}{s_1^2+1}\right)\biggl[\frac{d}{ds_3}\mathcal{F}_\delta(\infty,0;s_3,s_1)\biggl]_{s_3=s_1}+\nonumber\\
&+\frac 12\int_0^\infty ds_1 \int_0^{s_1} ds_3 
\log(s_1^2+1)\log\left(\frac{s_3^2}{s_3^2+1}\right)\frac{d}{ds_3}\frac{d}{ds_1}\mathcal{F}_\delta(\infty,0;s_3,s_1)\Biggl]
\end{align} 
and
 \begin{align}
 \label{intpartia}
&\text{(b)}+\text{(c)}=\frac{g_{2d}^6N^4}{32\pi^3}\Biggl[-\log^2\epsilon 
\biggl[\mathcal{F}_\delta(\infty,0;0,\infty)+\mathcal{F}_{-\delta}(\infty,0;0,\infty)\biggl]+\nonumber\\
&+\frac 12 \log\epsilon \int_\infty^0 dt_1
\log\left(\frac{t_1^2}{t_1^2+1}\right)\frac{d}{dt_1}
\biggl[\mathcal{F}_\delta(\infty,t_1;0,\infty)+\mathcal{F}_{-\delta}(\infty,t_1;0,\infty)\biggl]-\nonumber\\
&-\frac 12 \log \epsilon \int^\infty_0 ds_1
\log\left(s_1^2+1\right)\frac{d}{ds_1}
\biggl[\mathcal{F}_\delta(\infty,0;0,s_1)+\mathcal{F}_{-\delta}(\infty,0;0,s_1)\biggl]+\\
&
+\frac 14 \int_\infty^0 dt_1\int_0^\infty ds_1
\log\left(s_1^2+1\right)\log\left(\frac{t_1^2}{t_1^2+1}\right)\frac{d}{dt_1}\frac{d}{ds_1}
\biggl[\mathcal{F}_\delta(\infty,t_1;0,s_1)+\mathcal{F}_{-\delta}(\infty,t_1;0,s_1)\biggl]\,,\nonumber
\end{align} 
where $\epsilon$ is a regulator that cuts the contour in a neighborhood  of  
 the operator insertions and it will be send to zero at the end of the computation. Now using that
\begin{equation}\begin{split}
\frac{d}{d a}\mathcal{F}_{\pm\delta}(a,0;0,\infty)=\frac{d}{d 
a}\mathcal{F}_{\mp\delta}(\infty,0;0,a)\,,\ \ \ \ 
\frac{d}{d a}\mathcal{F}_{\pm\delta}(\infty,0;a,\infty)=\frac{d}{d 
a}\mathcal{F}_{\mp\delta}(\infty,a;0,\infty)
\end{split}\end{equation} 
and the definition \eqref{ObsYM2}, we can sum up all the contributions in \eqref{intparti} and \eqref{intpartia} 
obtaining
\begin{align}\label{somma2d2}
&\mathcal{M}_1^{(1)}=-\frac{g_{2d}^4N^2}{4\pi^2}\Biggl\{
\frac 12\int_0^\infty\!\!\!\! 
ds_1\log(s_1^2+1)\log\left(\frac{s_1^2}{s_1^2+1}\right)\biggl[\frac{d}{ds_3}\mathcal{F}_\delta(\infty,0;s_3,s_1)\biggl]_{s_3=s_1}\\
&-\frac 14 \int^\infty_0\!\!\!\!\! dt_1\int_0^\infty\!\!\!\!\! ds_1
\log\left(s_1^2+1\right)\log\left(\frac{t_1^2}{t_1^2+1}\right)\frac{d}{dt_1}\frac{d}{ds_1}
\biggl[\mathcal{F}_\delta(\infty,t_1;0,s_1)+\mathcal{F}_{-\delta}(\infty,t_1;0,s_1)\biggl]\!\nonumber
\Biggl\}\,,
\end{align}
where we have defined $\mathcal{M}_L^{(l)}$ as the sum of the $\delta$-dependent 
part of $\langle \mathcal{W}_L^{(2d)}(\delta)\rangle$ at loop order $l$.
To derive the above equation we have taken advantage of the useful relations 
\begin{equation}\begin{split}
&\mathcal{F}_\delta(\infty,0;0,\infty)=\mathcal{F}_{-\delta}(\infty,0;0,\infty)\,,\ \ \ \ \ \ \ \  
\frac{d}{da}\frac{d}{db}\mathcal{F}_{\pm\delta}(\infty,0;a,b)=0
\,,\\
& \int^\infty_0 da
\log\left(\frac{a^2}{a^2+1}\right)\frac{d}{da}
\mathcal{F}_{\pm\delta}(\infty,a;0,\infty)=\int^\infty_0 db
\log\left(b^2+1\right)\frac{d}{db}
\mathcal{F}_{\pm\delta}(\infty,0;0,b)\,.
\end{split}\end{equation}
We remark that the dependence on the cutoff $\epsilon$ disappears: as expected we end up with a finite result. 
We further observe that turning $\delta\rightarrow -\delta$ is the same as taking complex 
conjugation, then we can rewrite \eqref{somma2d2} as follows
\begin{equation}\begin{split}\label{somma2d3}
&\mathcal{M}_1^{(1)}=-\frac{g_{2d}^4N^2}{16\pi^2}\Biggl\{
\int_0^\infty 
ds_1\log(s_1^2+1)\log\left(\frac{s_1^2}{s_1^2+1}\right)\biggl[\frac{d}{ds_3}\mathcal{F}_\delta(\infty,0;s_3,s_1)\biggl]_{s_3=s_1}\\
&-\int^\infty_0 dt_1\int_0^\infty ds_1
\log\left(s_1^2+1\right)\log\left(\frac{t_1^2}{t_1^2+1}\right)\frac{d}{dt_1}\frac{d}{ds_1}
\mathcal{F}_\delta(\infty,t_1;0,s_1)
\Biggl\}+\text{c.c.}
\end{split}\end{equation} 
and performing the derivatives \eqref{somma2d3} becomes
\begin{equation}\begin{split}\label{somma2d4}
\mathcal{M}_1^{(1)}=-\frac{g_{2d}^4N^2}{16\pi^2}
\int_0^\infty\!\!\!\! ds dt
\frac{ \log(s^2+1)}{(t^2+1)(s^2+1)}\frac{e^{i\delta}t-s}{t-e^{i\delta}s}\biggl[
\log\left(\frac{s^2}{s^2+1}\right)-\log\left(\frac{t^2}{t^2+1}\right)\biggl]+\text{c.c.}
\end{split}\end{equation} 
The function $\mathcal{H}_1$ is then obtained with the help of  \eqref{defBHL2}, then at one-loop order we have
\begin{equation}\label{pigreco}
\mathcal{H}_1^{(1)}(\lambda,\varphi)=\left. \frac{1}{2}\partial_{\delta} \mathcal{M}_1^{(1)}\right|_{\mycom{\delta=\pi-\varphi 
}{g_{2d}^2=-{2g^2}/{A}}}\,.
\end{equation}
Exploiting the simple decomposition
\begin{equation}\label{scomposizione}
\frac{e^{i\delta}t-s}{t-e^{i\delta}s}=-\frac t s+\frac 1 s \frac{s^2+1}{e^{i\delta}s-t}
-\frac 1 s \frac{t^2+1}{e^{i\delta}s-t}\,,
\end{equation}
the derivative of \eqref{somma2d4} becomes
\small
\begin{equation}\begin{split}\label{somma2d5}
\frac{g_{2d}^4N^2}{16\pi^2}
\!\!\int_0^\infty \!\!\!\!\!\!ds dt \frac{\log(s^2+1)}{s}
\biggl[\frac{1}{(t^2+1)}\!-\!\frac{1}{(s^2+1)}\biggl]\biggl[
\log\left(\frac{s^2}{s^2+1}\right)\!-\!\log\left(\frac{t^2}{t^2+1}\right)\biggl]\partial_\delta\biggl(\frac{1}{e^{i\delta}s-t}\biggl)+\text{c.c.}
\end{split}\end{equation} 
\normalsize
Using
\begin{equation}\label{derry}
\partial_\delta\biggl(\frac{1}{e^{i\delta}s-t}\biggl)
=-is\,e^{i\delta}\partial_t\biggl(\frac{1}{e^{i\delta}s-t}\biggl)\,,
\end{equation}
the integral \eqref{somma2d5} takes the form:
\begin{align}\label{somma2d52}
&\frac{ig_{2d}^4N^2}{16\pi^2}
\!\!\int_0^\infty \!\!\!ds \biggl\{
\log(s^2+1)\log\left(\frac{s^2}{s^2+1}\right)
\frac{s}{s^2+1}-\\
&-2\!\!\int_0^\infty\!\!\!\! dt \biggl[\log\left(\frac{s^2}{s^2+1}\right)
\frac{t\log(s^2+1)}{(t^2+1)^2}\frac{e^{i\delta}}{e^{i\delta}s-t}-\log\left(\frac{t^2}{t^2+1}\right)
\frac{1-t^2}{1+t^2}\frac{s}{1+s^2}
\frac{1}{e^{i\delta}s-t} \biggl]\biggl\}+\text{c.c.}\nonumber
\end{align}
\normalsize
Summing up the integrands with their complex conjugate (the first term vanishes)
and changing variables $s\rightarrow \sqrt{\omega \rho}$ and $t\rightarrow \sqrt{\omega/\rho}$, 
we obtain:
\begin{align}\label{somma2d6}
\partial_{\delta} \mathcal{M}_1^{(1)}&= 
-\frac{g_{2d}^4N^2}{8\pi^2}
\!\!\int_0^\infty \!\!\!d\omega d\rho
\frac{\sin\delta \;[(\rho+\omega)(\rho\omega-1)+\rho(1+\rho\omega)\log(1+\rho\omega)]\log\left(\frac{\rho\omega}{1+\rho\omega}\right)}
{(1+\rho\omega)(\rho+\omega)^2(\rho^2-2\rho\cos\delta +1)}\nonumber\\
&=\frac{g_{2d}^4N^2}{4\pi^2}\sin\delta\,
\!\!\int_0^\infty \!\!\!d\rho
\frac{\log^2\rho}
{(\rho^2-1)(\rho^2-2\rho\cos\delta +1)}\, .
\end{align} 
The integration domain is easily restricted to $[0,1]$ and we arrive at
\begin{equation}\begin{split}\label{somma2d6}
\partial_{\delta} \mathcal{M}_1^{(1)}&=
-\frac{g_{2d}^4N^2}{4\pi^2}\sin\delta\,
\!\!\int_0^1 \!\!\!d\rho
\frac{\log^2\rho}
{\rho^2-2\rho\cos\delta +1}
=-g_{2d}^4N^2\,\frac{\pi}{3}\,
B_3\left(\frac{\delta}{2\pi}\right),
\end{split}\end{equation} 
where the $B_n(x)$ are the Bernoulli polynomials defined as
\begin{equation}
\label{Bernoulli}
B_{2n+1}(x)=\frac{(-1)^{n+1}2(2n+1)\sin(2\pi x)}{(2\pi)^{2n+1}}
\int_0^1 dt \;\frac{\log^{2n} t}{t^2-2t\cos(2\pi x)+1}.
\end{equation} 
Finally, using \eqref{pigreco}, we find the desired result
\begin{equation}\mathcal{H}_1^{(1)}(\lambda,\varphi)=
-\frac{\lambda^2}{24 \pi}\,
B_3\left(\frac{\pi-\varphi}{2\pi}\right).
\end{equation} 

\subsection{Operator insertions of length $L$}\label{sec:casoL}

 \begin{figure}[ht]
\centering
	\includegraphics[width=12cm]{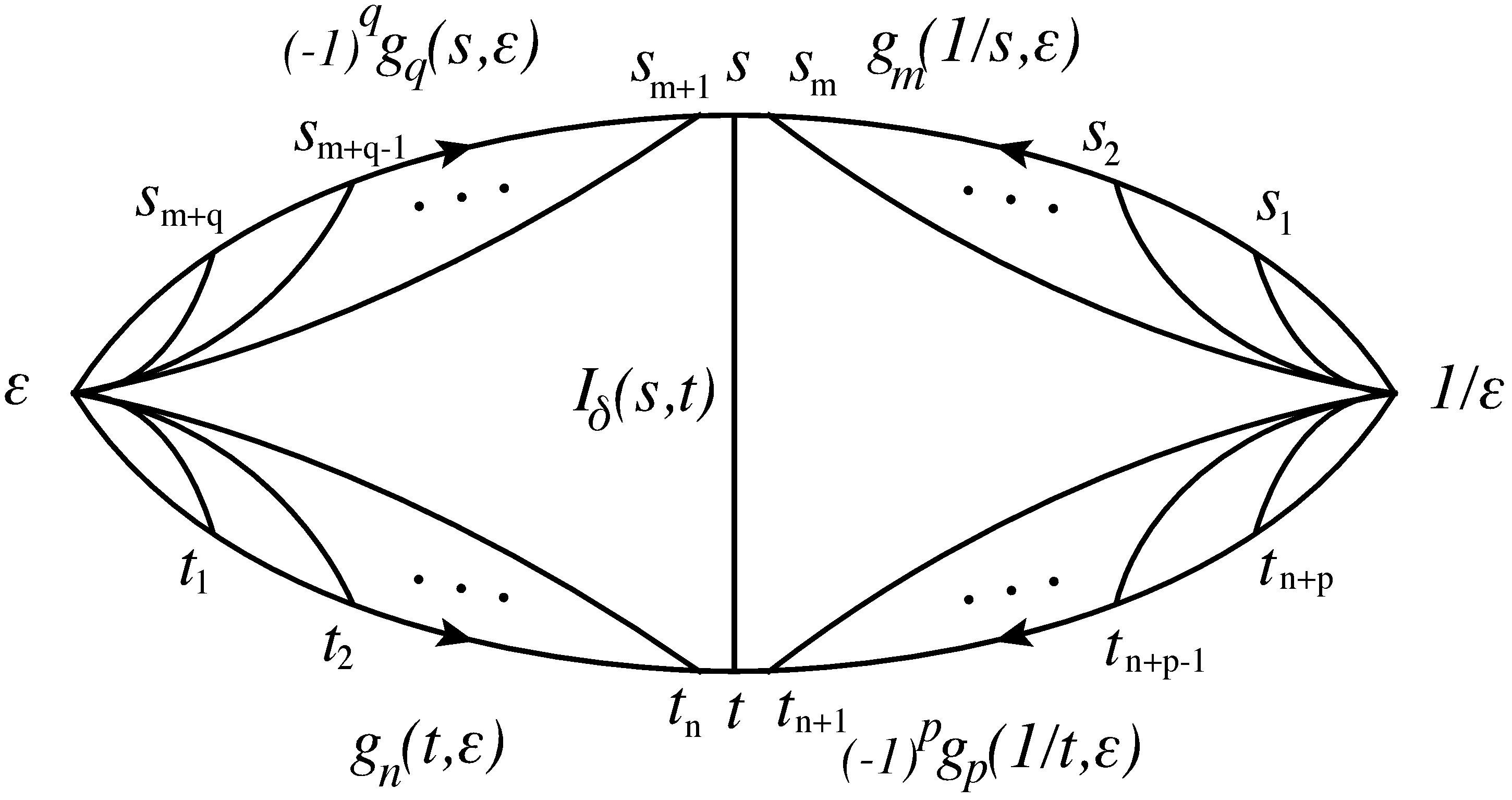}
	\caption{Schematic representation of an arbitrary diagram contributing to the first non-trivial order of $\mathcal{B}_L(\lambda,\varphi)$}
	\label{fig:spicchiogenerale}
\end{figure}

We are now ready to compute the leading weak coupling contribution to the L\"{u}scher term, using YM$_2$ perturbation theory. We have to consider all possible planar diagrams with one single line connecting the right and left sides of the wedge, in presence of length $L$ operator insertions (see Figure \ref{fig:spicchiogenerale}). 
As remarked in the previous Section, every diagram has the same weight.

We construct  the generic term of the relevant perturbative order by introducing the 
auxiliary function $g_n$, 
that contains the contribution of all the propagators connecting the operators with contour
in each of the four sectors represented in Figure \ref{fig:spicchiogenerale}:
\begin{equation}\label{emmeelle1}
\mathcal{M}_L^{(1)}\!=\frac{(-1)^{L+1}2^{3L}\,\pi^L}{g_{2d}^{\;\,2L}\, N^{L}}
\!\!\sum_{n,m=0}^L\!\int_{0}^\infty \!\!\!\!\!ds 
dt\,(-1)^{n+m}g_n(t,\epsilon)g_{L-n}(s,\epsilon)I_\delta(s,t)g_m(1/s,\epsilon)g_{L-m}(1/t,\epsilon)
\end{equation}
where 
\begin{equation}\label{idelta}
I_{\delta}(s,t)=\left(-\frac{g_{2d}^2}{2\pi}\right)\frac{1}{(t^2+1)(s^2+1)}\frac{t e^{i\delta}-s}{t-e^{i\delta}s}
\end{equation}
and
\begin{equation}\label{gdef}
g_n(t,\epsilon)=
\left(-\frac{g_{2d}^2\,N}{4\pi}\right)^n\int_\epsilon^t 
dt_n\int_\epsilon^{t_n} dt_{n-1} \;...\int_\epsilon^{t_2}dt_1\; \prod_{i=1}^n\Delta(t_i),
\end{equation}
being $\Delta(t)=\frac 1 t\frac{1}{t^2+1}$ the propagator from the origin to the contour
(up to a constant factor, see \eqref{prop2dreal}). 
Obviously the number of lines on the right and on the left side of the central propagator is equal to $L$. Moreover we have introduced an explicit cutoff
$\epsilon\rightarrow 0$ to avoid intermediate divergencies. 
A recurrence relation for $g_n$ and its derivative follows directly from its definition:
\begin{equation}\begin{split}\label{recrel}
g_n(t,\epsilon)=&\left(-\frac{g_{2d}^2\,N}{4\pi}\right)\int_\epsilon^t 
dt_n\,\Delta(t_n)\,g_{n-1}(t_n,\epsilon)\,,\\
\frac{d}{dt}g_n(t,\epsilon)=&\left(-\frac{g_{2d}^2\,N}{4\pi}\right)\,\Delta(t)\,g_{n-1}(t,\epsilon)\,.
\end{split}\end{equation}
As shown in the Appendix \ref{sec:appC}, we can combine these two equations into a single recurrence relation, involving just $g_n(t,\epsilon)$
\begin{equation}\label{recurrvera}
g_n(t,\epsilon)=-\sum_{k=1}^n\frac{(-\alpha)^k}{k!}\log^k\left(\frac{t^2}{t^2+1}\right)\,
g_{n-k}(t,\epsilon)-(t\rightarrow \epsilon),
\end{equation} 
where $\alpha=-\frac{g_{2d}^2\,N}{8\pi}$. We can solve the recurrence relation by finding the related generating function $G(t,\epsilon,z)$
\begin{equation}
g_n(t,\epsilon)=\frac{1}{2\pi i}\oint_\gamma 
\frac{dz}{z^{n+1}}G(t,\epsilon,z)\,,
\end{equation}
for a suitable closed curve $\gamma$ around the origin. We have obtained (see Appendix \ref{sec:appC} for the details)
\begin{equation}\label{defg}
G(t,\epsilon,z)=\left(\frac{t^2}{t^2+1}\frac{\epsilon^2+1}{\epsilon^2}\right)^{\alpha 
z}\,.
\end{equation}
Then \eqref{emmeelle1} becomes
\small\begin{equation}\label{Wres}
\mathcal{M}_L^{(1)}=\frac{(-1)^{L+1}}{(2\pi i)^4}\frac{2^{3L}\,\pi^L}{g_{2d}^{\;\,2L}\, N^{L}}
\sum_{n,m=0}^L\int_{0}^\infty\!\!\!\!\! ds 
dt\,\oint \frac{G(t,\epsilon,z)}{z^{n+1}}\frac{G^{-1}(s,\epsilon,w)}{w^{L-n+1}}
I_\delta(s,t)\frac{G(1/s,\epsilon,v)}{v^{m+1}}\frac{G^{-1}(1/t,\epsilon,u)}{u^{L-m+1}}.
\end{equation}\normalsize
The sums over $m$ and $n$ are done explicitly
\begin{equation}
\label{geosum}
\sum_{n=0}^L\frac{1}{z^{n+1}w^{L-n+1}}=\frac{z^{-(L+1)}-w^{-(L+1)}}{w-z}
\end{equation}
and using the observations in Appendix \ref{sec:appC} and in particular the 
formula \eqref{residui42}, we obtain
\small\begin{equation}\begin{split}\label{Wres3}
\mathcal{M}_L^{(1)}&=\frac{(-1)^{L+1}}{(2\pi i)^2}\frac{2^{3L}\,\pi^L}{g_{2d}^{\;\,2L}\, N^{L}}
\left(\frac{-g_{2d}^2}{2\pi}\right)\times\\
&\times\int_{0}^\infty\!\!\! ds 
dt\!\oint \frac{dz}{z^{L+1}} \frac{dv}{v^{L+1}}
\biggl[\left(\frac{t^2(1+s^2)}{s^2(1+t^2)}\right)^{\alpha z}\!\!
\left(\frac{(1+t^2)}{(1+s^2)}\right)^{\alpha v}\!\!\!\!\frac{1}{(t^2+1)(s^2+1)}
\frac{t e^{i\delta}-s}{t-e^{i\delta}s}\biggl]\,.
\end{split}\end{equation}\normalsize
Notice that the $\epsilon$-dependence has disappeared: the cancellation of the intermediate 
divergencies is a non-trivial bonus of our method.

\noindent
Computing the residues, \eqref{Wres3} takes the following compact form
\begin{equation}\begin{split}\label{Wres4}
\mathcal{M}_L^{(1)}=\frac{(-1)^{L+1}\alpha^{2L}}{(L!)^2}\frac{2^{3L}\,\pi^L}{g_{2d}^{\;\,2L}\, N^{L}}
\left(\frac{-g_{2d}^2}{2\pi}\right)\int_{0}^\infty\!\!\! ds dt
\frac{\left[\log\left(\frac{t^2(1+s^2)}{s^2(1+t^2)}\right)\log\left(\frac{(1+t^2)}{(1+s^2)}\right)\right]^L}{(t^2+1)(s^2+1)}
\frac{t e^{i\delta}-s}{t-e^{i\delta}s}\,.
\end{split}\end{equation}
We are interested in calculating the Bremsstrahlung function, so we take the derivative 
of the VEV of the Wilson loop as seen in the equation \eqref{defBH}.
Using (\ref{derry}), after some algebra, we find
\begin{align}\label{Wres6}
\mathcal{H}_L^{(1)}=i\frac{(-1)^{L+1}\alpha^{2L}}{2(L!)^2}&\frac{2^{3L}\,\pi^L}{g_{2d}^{\;\,2L}\, N^{L}}
\left(\frac{-g_{2d}^2}{2\pi}\right)\int_{0}^\infty\!\!\! ds dt
\Biggl\{\biggl[\log\left(\frac{t^2(1+s^2)}{s^2(1+t^2)}\right)\log\left(\frac{(1+t^2)}{(1+s^2)}\right)\biggl]^L\nonumber\\
&\left.\times\left[\frac{1}{t^2+1}\partial_{s}\left(\frac{1}{e^{i\delta}s-t}\right)
+\frac{e^{i\delta}}{s^2+1}\partial_{t}\left(\frac{1}{e^{i\delta}s-t}\right)\right]\biggr\}\right|_{\mycom{\delta=\pi-\varphi 
}{g_{2d}^2=-{2g^2}/{A}}}\,.
\end{align}
Performing some integration by parts and transformations  $s\leftrightarrow t$ and $s,t\rightarrow \frac{1}{s,t}$,
the change of variables $s\rightarrow \sqrt{\omega\rho}$ and $t\rightarrow \sqrt{\omega/\rho}$ gives
\begin{align}\label{quasifinito}
&\mathcal{H}_L^{(1)}=-\frac{(-1)^{L+1}\alpha^{2L}}{L!(L-1)!}\frac{2^{3L}\,\pi^L}{g_{2d}^{\;\,2L}\, N^{L}}
\left(\frac{-g_{2d}^2}{2\pi}\right)\sin\delta\\ &\times\!\!\! \int_{0}^1\!\!\! \!\!d\rho\!\int_{0}^\infty\!\!\!\!\!\! d\omega\!\!
\left.\biggl[\log\!\left(\frac{\rho(\rho+\omega)}{\rho\omega+1}\right)\!\log\!\left(\frac{\rho(\rho\omega+1)}{\rho+\omega}\right)\!\!\biggl]^{L-1}
\!\!\!\!\!\!\!\!\!\!
\frac{(\rho^2-1)\log\rho}{(\rho+\omega)(1+\rho\omega)(\rho^2-2\rho\cos\delta +1)}
\right|_{\mycom{\delta=\pi-\varphi }{g_{2d}^2=-{2g^2}/{A}}}.\nonumber
\end{align}
Using the expansion
\begin{equation}
\biggl[\log\left(\frac{\rho(\rho+\omega)}{\rho\omega+1}\right)\log\left(\frac{\rho(\rho\omega+1)}{\rho+\omega}\right)\biggl]^{L-1}\!\!
=\sum_{k=0}^{L-1}\binom{L-1}{k}(-1)^k\log^{2(L-1-k)}\rho\log^{2k}\left(\frac{\rho+\omega}{\rho\omega+1}\right)\,,
\end{equation}
the integration over $\omega$ is straightforward since 
\begin{equation}
\int d\omega 
\frac{\log^{2k}\left(\frac{\rho+\omega}{\rho\omega+1}\right)}{(\rho+\omega)(1+\rho\omega)}=
-\frac{\log^{2k+1}\left(\frac{\rho+\omega}{\rho\omega+1}\right)}{(\rho^2-1)(2k+1)}\,.
\end{equation}
Then we obtain
\small\begin{equation}\label{penultimo}
\left.
\mathcal{H}_L^{(1)}=-\frac{(-1)^{L+1}\alpha^{2L}}{L!(L-1)!}\frac{2^{3L}\,\pi^L}{g_{2d}^{\;\,2L}\, N^{L}}
\left(\frac{-g_{2d}^2}{2\pi}\right)
\beta\left(\frac 12,L\right)
\biggl\{\sin\delta \int_{0}^1\!\!\! d\rho
\frac{\log^{2L}\rho}{\rho^2-2\rho\cos\delta+1}
\Biggl\}\right|_{\mycom{\delta=\pi-\varphi }{g_{2d}^2=-{2g^2}/{A}}}\,,
\end{equation}\normalsize
where $\beta(a,b)$ is the Euler Beta function.
The integral in \eqref{penultimo} is basically the standard representation \eqref{Bernoulli} of the Bernoulli polynomials.
Therefore $\mathcal{H}_L^{(1)}$ becomes:
\begin{equation}\begin{split}
\mathcal{H}_L^{(1)}=-\frac{(-1)^{L+1}\alpha^{2L}}{L!(L-1)!}\frac{2^{3L}\,\pi^L}{g_{2d}^{\;\,2L}\, N^{L}}
\left(\frac{-g_{2d}^2}{2\pi}\right)
\beta\left(\frac 12,L\right)
\frac{(2\pi)^{2L+1}(-1)^{L+1}}{2(2L+1)}B_{2L+1}\left(\frac{\pi-\varphi}{2\pi}\right)\,.
\end{split}\end{equation}
Finally inserting the expression for $\alpha$ and using
 \eqref{coup2D4D}, we obtain the desired result
\begin{equation}\begin{split}
\mathcal{H}_L^{(1)}=-\frac{(-1)^{L}\lambda^{L+1}}{4\pi(2L+1)!}B_{2L+1}\left(\frac{\pi-\varphi}{2\pi}\right).
\end{split}\end{equation}

\section{The $L=1$ case from perturbative $\mathcal{N}=4$ SYM}

 \begin{figure}[!h]
 \centering
  \subfigure[$H_1$]
   {\includegraphics[width=2.7cm]{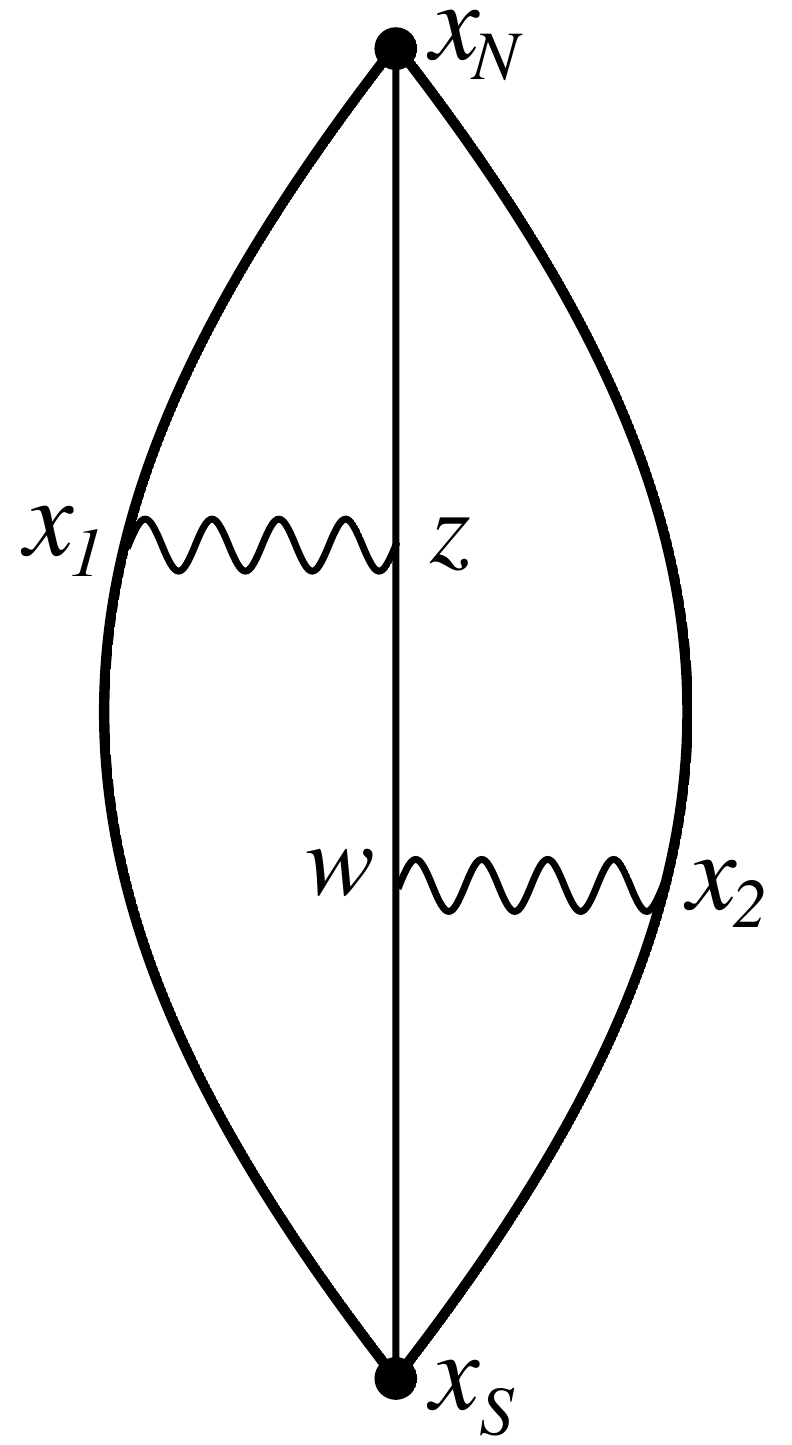}}
    \hspace{5mm}
    \subfigure[$H_2$]
   {\includegraphics[width=2.7cm]{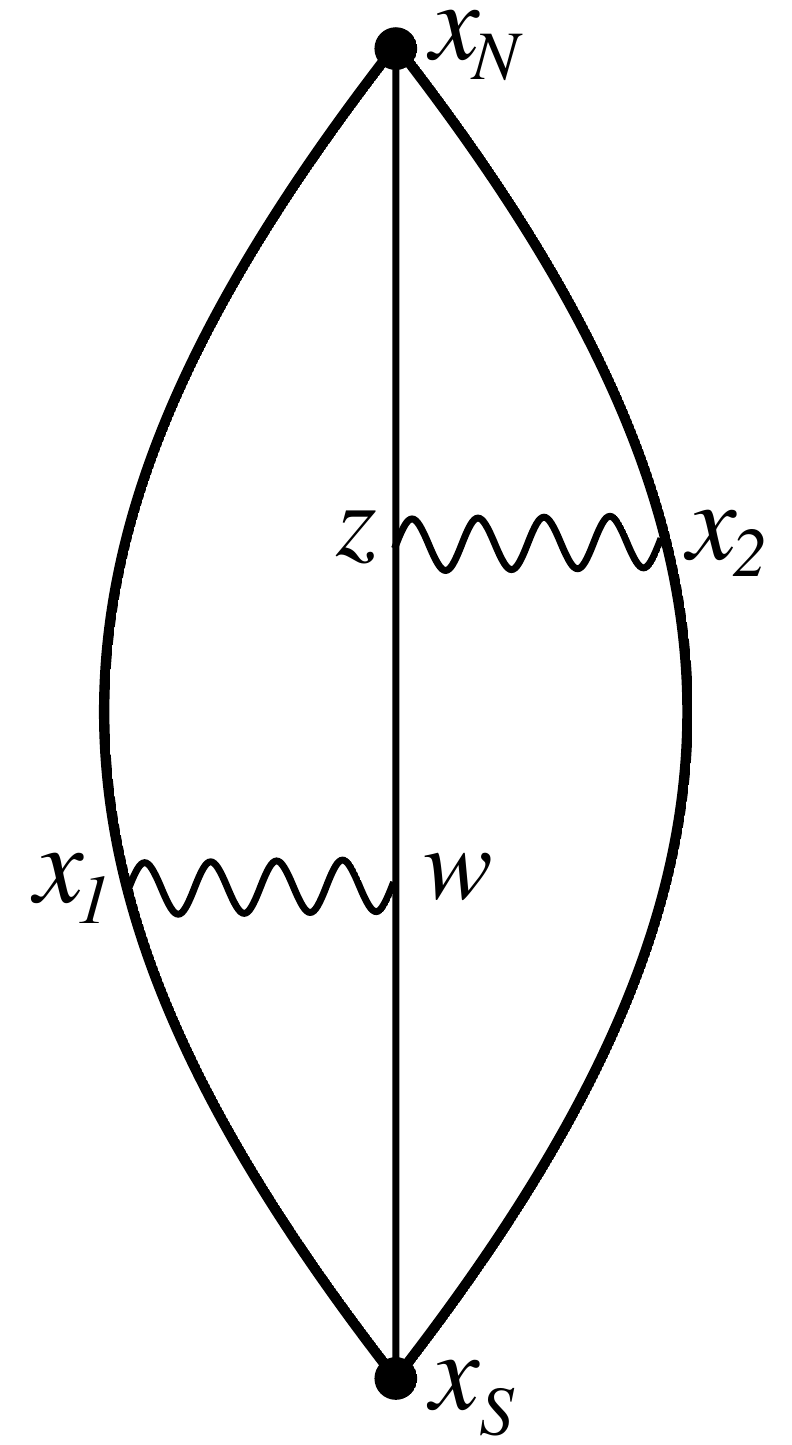}}
    \hspace{5mm}
 \subfigure[$X_1$]
   {\includegraphics[width=2.7cm]{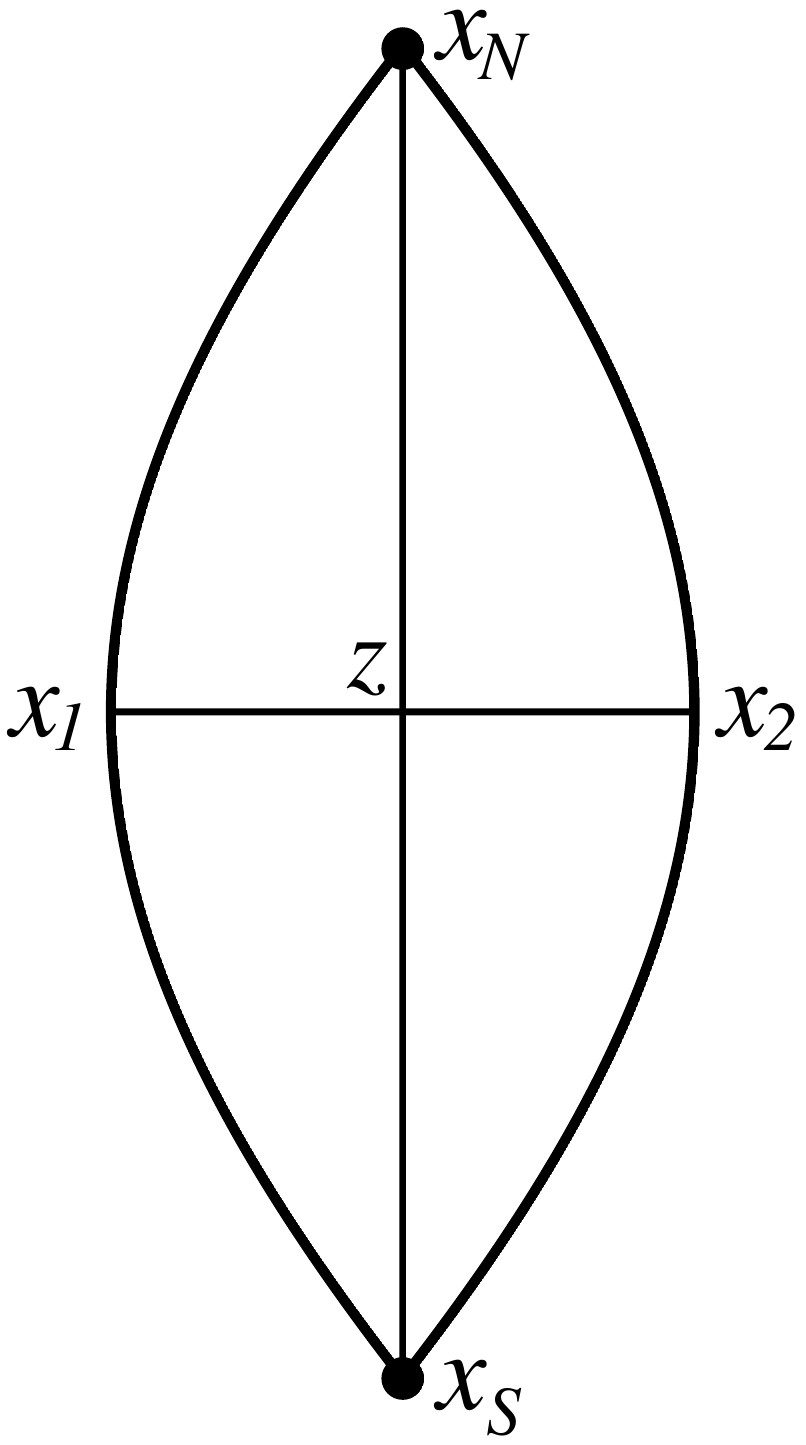}}
        \hspace{5mm}
 \subfigure[$X_2$]
   {\includegraphics[width=2.7cm]{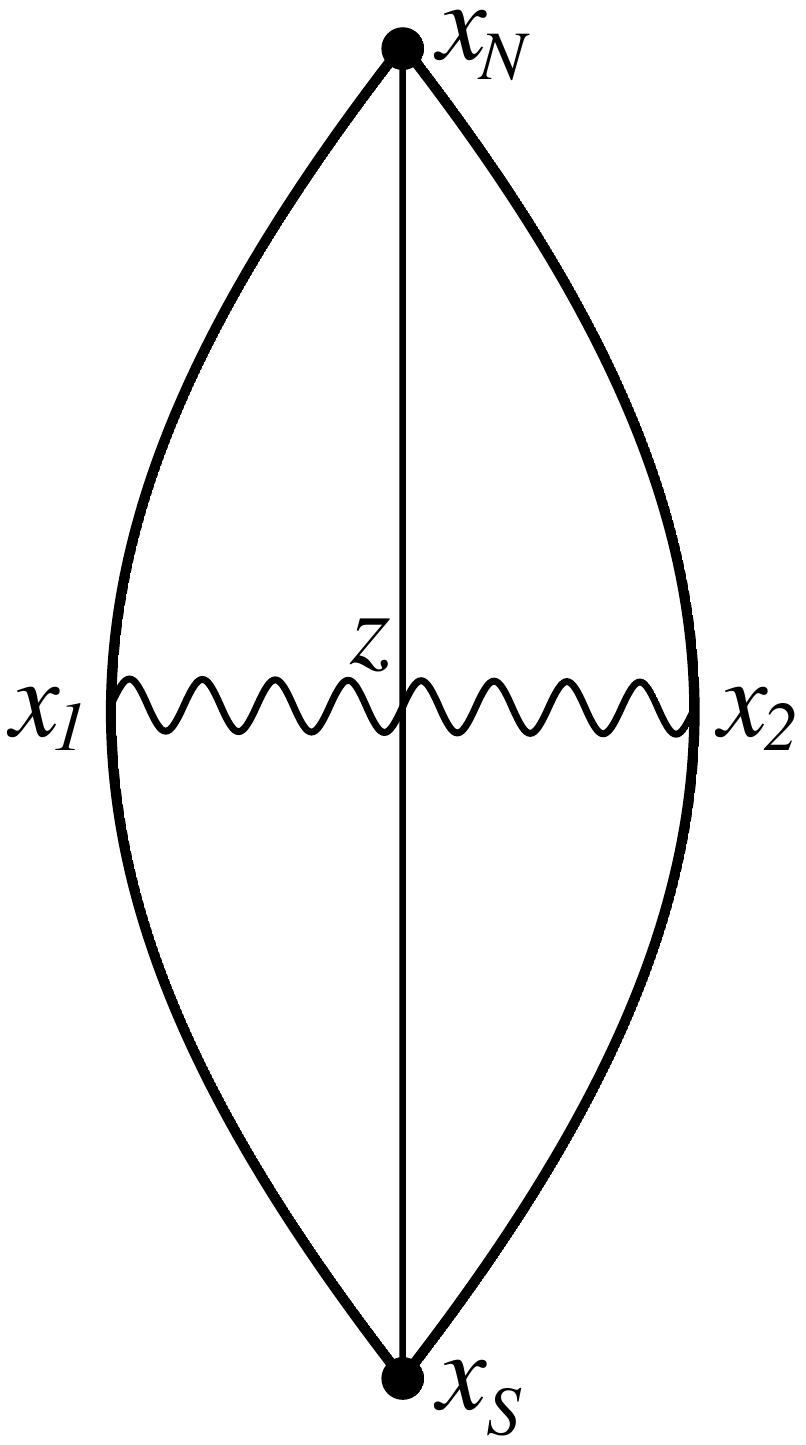}}
    \hspace{5mm}
 \caption{All the $\delta$-dependent diagrams at order $\lambda^2$}
  \label{fig:3grafici}
 \end{figure}

In order to check the above results, in this Section we compute the one-loop Feynman diagrams associated to the Bremsstrahlung 
function for the operator insertion with $L=1$. It can be considered also as further test of the 
correspondence between $\mathcal{N}=4$ BPS observables on $S^2$ and two-dimensional 
Yang-Mills theory on the sphere. We are interested only in the $\delta$-dependent contributions, 
then we have to consider just diagrams which connect the right and left sides of the wedge, at 
the first non-trivial order. In Figure \ref{fig:3grafici} we represent all the $\delta$-dependent 
Feynman graphs we can draw at order $\lambda^2$.

The diagram (a) of Figure \ref{fig:3grafici} is given by
\begin{equation}\begin{split}\label{H1inizio}
[3.(\text{a})]\!=\!\frac{\lambda^2}{4}
\!\int_0^\pi \!\!\! d\tau_1\!\!\int_\pi^{2\pi}\!\!\!\! d\tau_2\,
\dot{x}_1^\mu \dot{x}_2^\nu \left(\partial_{x_1^\mu}\partial_{x_2^\nu}\!+\!
2\partial_{x_1^\mu}\partial_{x_S^\nu}\!+\!2\partial_{x_N^\mu}\partial_{x_2^\nu}
\!+\!4\partial_{x_N^\mu}\partial_{x_S^\nu}\right)\!\mathcal{H}(x_1,x_N;x_2,x_S)\,,
\end{split}\end{equation}
where
\begin{equation}
\mathcal{H}(x_1,x_2;x_3,x_4)\equiv \int \! \frac{d^4 z\, d^4 w}{(2\pi)^8}
D(x_1-z)D(x_2-z)D(w-z)D(x_3-w)D(x_4-w)
\end{equation}
with $D(x-y)=1/(x-y)^2$. All but the last term in \eqref{H1inizio} contain at 
least one derivative with respect to $\tau_1$ or $\tau_2$: these terms, after 
the integration over $\tau$, are $\delta$-independent.
Therefore we are left with
\begin{equation}\begin{split}\label{int2}
H_1&=\lambda^2\!\!
\int_0^\pi \!\!\! d\tau_1\int_\pi^{2\pi}\!\!\!\! d\tau_2\,
\dot{x}_1^\mu \dot{x}_2^\nu 
\,\partial_{x_N^\mu}\partial_{x_S^\nu}\mathcal{H}(x_1,x_N;x_2,x_S)\\
&=\!\frac{\lambda^2}{(2\pi)^2}\!\!
\int_0^\pi \!\!\!\! d\tau_1\!\!\int_\pi^{2\pi}\!\!\!\!\! d\tau_2\!\!\int\! d^4 z \,
(\dot{x}_1\!\cdot\!\partial_{x_N})\,D(x_1\!-\!z)D(x_N\!-\!z)
(\dot{x}_2\!\cdot\! \partial_{x_S})\,\mathcal{I}_1(x_S\!-\!z,x_2\!-\!z)\,,
\end{split}\end{equation}
where
 \begin{equation}\label{I1sopra}
\mathcal{I}_1(x_1-x_3,x_2-x_3)\equiv \frac{1}{(2\pi)^6}\int d^4 w\, 
D(x_1-w)D(x_2-w)D(x_3-w)\,.
\end{equation} 
To compute one of the Feynman integrals in \eqref{int2}, we apply the trick 
used in \cite{Bassetto:2008yf}. In particular we add to the integrand a term which becomes $\delta$-independent 
after the $\tau$ integration
\begin{equation}
(\dot{x}_1\!\cdot\!\partial_{x_N})\,D(x_1\!-\!z)D(x_N\!-\!z)
(\dot{x}_2\!\cdot\! \partial_{x_2})\,\mathcal{I}_2(x_S\!-\!z,x_2\!-\!z)\,,
\end{equation}
where $\mathcal{I}_2$ is defined in \cite{Bassetto:2008yf}.
Therefore we can recast the $\delta$-dependence of \eqref{int2} as
\begin{equation}\label{integrandovumu}
\frac{\lambda^2}{(2\pi)^2}
\int_0^\pi \!\!\! d\tau_1\!\!\int_\pi^{2\pi}\!\!\!\! d\tau_2\!\!\int d^4 z \,
(\dot{x}_1\cdot\partial_{x_N})\,D(x_1\!-\!z)D(x_N\!-\!z)\;
\dot{x}_2\cdot V(x_S\!-\!z,x_2\!-\!z)\,,
\end{equation}
where
\begin{equation}\label{vumudef}
V^\mu(x,y)\equiv \partial_{x^\mu}\mathcal{I}_1({x},{y})-
\partial_{y^\mu} \mathcal{I}_2({x},{y}).
\end{equation}
Using the explicit representation for $V^\mu$ (see \cite{Bassetto:2008yf}), we can write
\small\begin{equation}
\begin{split}
\dot{x}_2\cdot & V(x_S\!-\!z,x_2\!-\!z)=\frac{D(x_S-z)}{32\pi^4}
\biggr\{\frac{d}{d\tau_2}\biggr[\mathrm{Li}_2
 \left(1-\frac{(x_S-x_2)^2}{(x_2-z)^2}\right)+\frac 12 
 \log^2\left(\frac{(x_S-x_2)^2}{(x_2-z)^2}\right)\\
 &\qquad\qquad\qquad-\frac 12 
 \log^2\left(\frac{(x_S-x_2)^2}{(x_S-z)^2}\right)\biggr]-
 2\dot{x}_2\cdot (x_S-x_2)D(x_S-x_2)\log\left(\frac{(x_2-z)^2}{(x_S-z)^2}\right)\biggr\}\,.
\end{split}\end{equation}\normalsize
Again the total derivative gives a $\delta$-independent contribution, then we have
\begin{equation}\begin{split}
H_1=-\frac{\lambda^2}{(2\pi)^6}
\int_0^\pi \!\!\! d\tau_1\!\int_\pi^{2\pi}\!\!\!\! d\tau_2& \,(\dot{x}_2\cdot x_S)D(x_S-x_2)\int \! d^4 z \,
(\dot{x}_1\cdot\partial_{x_N})\,D(x_1-z)\\
&D(x_N-z)D(x_S-z)\,
\log\!\left(\frac{(x_2-z)^2}{(x_S-z)^2}\right)+\text{``$\delta$-ind. terms''}\,.
\end{split}\end{equation}
The $\delta$-dependent part of the diagram in Figure \ref{fig:3grafici}.(b) is easily obtained from $H_1$ 
by exchanging $x_N\leftrightarrow x_S$.

\begin{figure}[ht]
\centering
	\includegraphics[width=10cm]{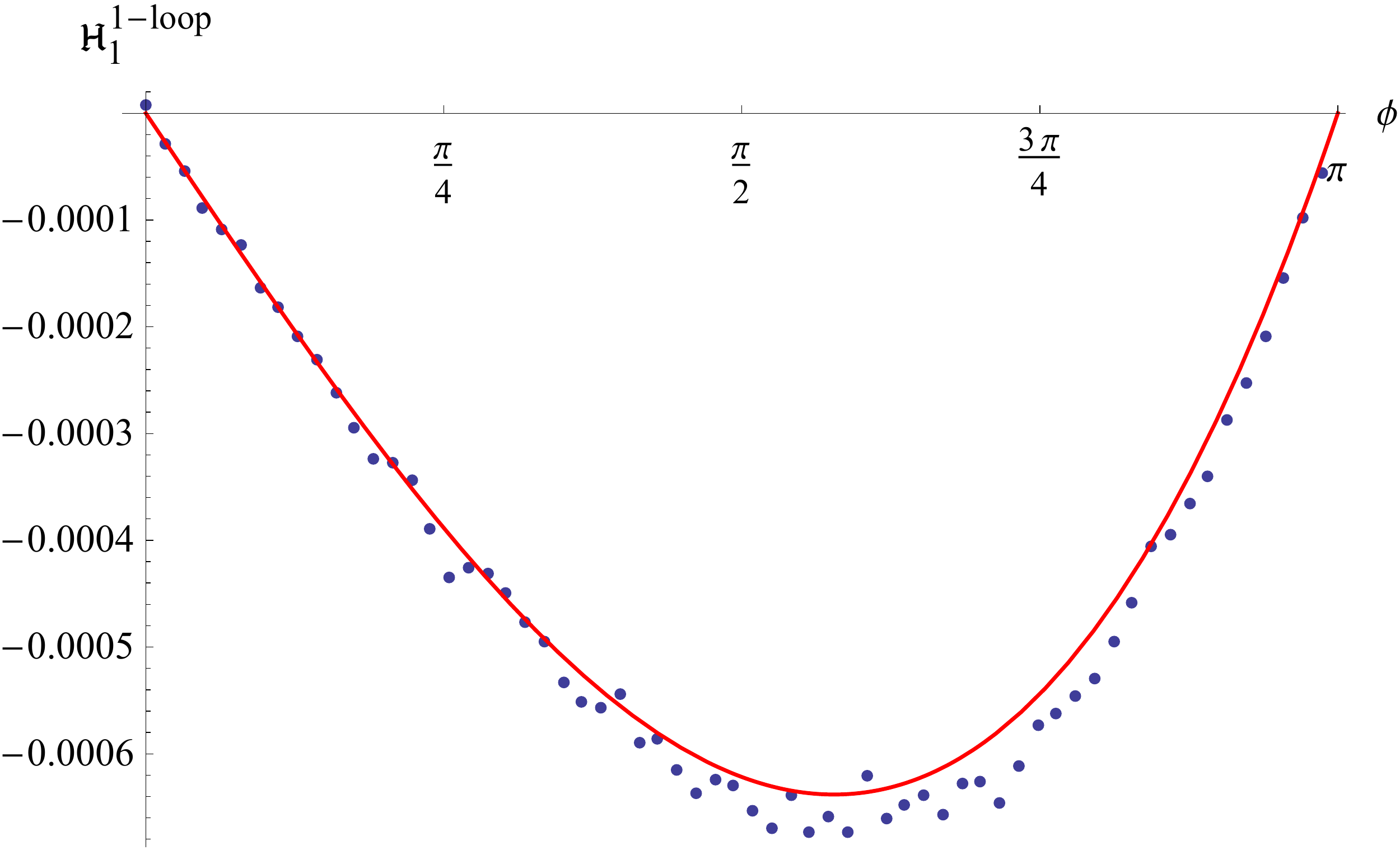}
		\caption{Numerical computation of $\mathcal{H}_1^{(1)}$. The blue dots are the numerical data and the red line is the expected curve.}
	\label{fig:numerico}
\end{figure}

The diagrams (c) and (d) of Figure \ref{fig:3grafici} are given by
\begin{equation}\begin{split}
[3.(c)]=X_1=&-\frac{\lambda^2}{4}\cos{\delta}\int_0^\pi d\tau_1\int_\pi^{2\pi}d\tau_2\;\mathcal{X}(1,1,1,1)\,,\\
[3.(d)]=X_2=&-\frac{\lambda^2}{4}\int_0^\pi d\tau_1\int_\pi^{2\pi}d\tau_2
\;\dot{x}_1\cdot\dot{x}_2 \;\mathcal{X}(1,1,1,1)\,,
\end{split}\end{equation}
where $\mathcal{X}$ is the scalar component of a more general class of tensorial 
Feynman integrals $\mathcal{X}^{\mu_1...\mu_n}$ defined as follows
\begin{equation}
\mathcal{X}^{\mu_1...\mu_n}(\nu_1,\nu_2,\nu_3,\nu_4)\equiv \!\!
\int \! \frac{d^4 z}{(2\pi)^6}\; 
z^{\mu_1}...z^{\mu_n}\,D(x_S\!-\!z)^{\nu_1}D(x_1\!-\!z)^{\nu_2}D(x_N\!-\!z)^{\nu_3}D(x_2\!-\!z)^{\nu_4}.
\end{equation}
Now, recalling the definition \eqref{defBHL} and denoting with the prime the derivative respect to $\delta$ (notice that $x_2$, $x_2'$ and $\dot{x}_2$ form an orthogonal basis), we have
\begin{equation}\begin{split}
H_1'=&4\lambda^2
\int_0^\pi \!\!\! d\tau_1\!\int_\pi^{2\pi}\!\!\!\! d\tau_2 \,(\dot{x}_2\cdot x_S)D(x_S-x_2)
\,x_2'^\mu\,\dot{x}_1^\nu\;\left(x_N^\nu\mathcal{X}^\mu(1,1,2,1)-
\mathcal{X}^{\mu\nu}(1,1,2,1)\right)\,,\\
H_2'=&4\lambda^2
\int_0^\pi \!\!\! d\tau_1\!\int_\pi^{2\pi}\!\!\!\! d\tau_2 \,(\dot{x}_2\cdot x_N)D(x_N-x_2)
\,x_2'^\mu\,\dot{x}_1^\nu\;\left(x_S^\nu\mathcal{X}^\mu(2,1,1,1)-
\mathcal{X}^{\mu\nu}(2,1,1,1)\right)\,,\\
X_1'=&\frac{\lambda^2}{4}\int_0^\pi d\tau_1\int_\pi^{2\pi}d\tau_2\;
\left(\sin{\delta}\,\mathcal{X}(1,1,1,1)-2\cos{\delta}\,x_2'^\mu\,\mathcal{X}^\mu(1,1,1,2)\right)\,,\\
X_2'=&\frac{\lambda^2}{4}\int_0^\pi d\tau_1\int_\pi^{2\pi}d\tau_2\;
\left(-(\dot{x}_1\cdot x_2')\,\mathcal{X}(1,1,1,1)+2(\dot{x}_1\cdot 
\dot{x}_2)\,x_2'^\mu\,\mathcal{X}^\mu(1,1,1,2)\right)\,.
\end{split}\end{equation}
Finally, summing up all the different contributions in ${\cal N}=4$ SYM, we obtain the 
function $\mathcal{H}_1^{(1)}$ defined in \eqref{defBHL}
\begin{equation}\label{luscherfine}
\mathcal{H}_1^{(1)}
=\left.\frac12 (H_1'+H_2'+X_1'+X_2')\;\right|_{\delta=\pi-\varphi}\,.
\end{equation}
In order to check the results of the previous sections, it is enough to perform
the integrals numerically. The result of the numerical computation is shown in Figure \ref{fig:numerico}. 

The numerical data are the average of several Montecarlo integrations. 
The data error is bigger when $\varphi$ is closed to $\pi/2$ because the 
integrals in the \eqref{luscherfine} are oscillatory.

In principle we could also compute analytically  the integrals in \eqref{luscherfine} with the help 
of usual Feynman integral techniques. Indeed, in the dual conformal symmetry picture, the quantities $\mathcal{X}$ are 
the so-called ``\textit{box}'' integrals.
Decomposing tensorial boxes in combinations of scalar ones, one should 
expand the result on a basis of the Master Integrals. Finally 
the tricky part would consist in computing the remaining integrals over the loop parameters. These quite technical computations are beyond the aim of this paper 
and we leave them to further developments in a may be more general setting.

\section{Conclusions and outlook}
In this paper we have explored the possibility to study the near-BPS expansion of the 
generalized cusp anomalous dimension with $L$ units of R-charge by means of 
supersymmetric localization. The R-charge is provided by the insertion certain scalar 
operators into a cusped Wilson loop, according to the original proposal of 
\cite{Correa:2012hh}, \cite{Drukker:2012de}. The relevant generalized Bremsstrahlung 
function $\mathcal{B}_L(\lambda,\varphi)$ has been computed by solving a set of TBA equations 
in the near-BPS limit \cite{Gromov:2012eu,Gromov:2013qga} and, more recently, using QSC 
approach \cite{Gromov:2015dfa}. We have proposed here a generalization of the method 
discussed in \cite{Correa:2012at}, relating the computation to the quantum average of some 
BPS Wilson loops with local operator insertions along the contour. The system should localize 
into perturbative YM$_2$ on $S^2$, in the zero-instanton sector, suggesting the possibility 
to perform exact calculations in this framework. We have checked our proposal, reproducing 
the leading L\"{u}scher correction at weak coupling to the generalized cusp anomalous 
dimension. We have further tested our strategy in the case $L=1$, using Feynman diagrams 
directly in ${\cal N} = 4$ Super Yang-Mills theory.

\noindent
Our investigations represent only a first step in connecting integrability results with 
localization outputs: we certainly would like to derive the complete expression for 
$\mathcal{B}_L(\lambda,\varphi)$ in this framework. Two-dimensional Yang-Mills theory on the 
sphere has an exact solution, even at finite $N$ \cite{Migdal:1975zf,Witten:1991we}: on the 
other hand, the construction of the vacuum expectation values of Wilson loops with local 
operator insertions has not been studied in the past, at least to our knowledge. 
The matrix model \cite{Sizov:2013joa}, computing the generalized Bremsstrahlung function, strongly 
suggests that these two-dimensional observables, in the zero-instanton sector, should be 
obtained by extending the techniques of \cite{Giombi:2012ep}. One could expect that 
also a finite $N$ answer is possible, as in the case of $L=0$. We are currently working on 
these topics and we hope to report some progress soon.

\noindent
A further direction could be to develop an efficient technique to explore this kind of 
observable directly in four-dimensions, by using perturbation theory. It would be interesting to 
go beyond the near-BPS case and to study the anomalous dimensions for more general 
local operator insertions\footnote{We thank Nadav Drukker for suggesting us these possibilities.}. 
The construction of similar systems in three-dimensional ABJM theory should also be 
feasible and could provide new insights to get exact results.

\section*{Acknowledgements}
We warmly thank N. Drukker and S. Giombi for  discussions and  suggestions. One of us  (M.P.) would like to thank  to  the Group "Gauge fields from strings"  at Humboldt University for the warm hospitality while this work has been completed.
This work has been supported in part by MIUR, INFN and MPNS-COST Action MP1210 ``The String Theory Universe''.

\appendix
\begin{flushleft}
\Large \bf Appendix
\end{flushleft}
\addcontentsline{toc}{section}{Appendix}
\section{The generating function $G(t,\epsilon,z)$ and some related properties}\label{sec:appC}
We present here the derivation of some results that have been used in computing the L\"{u}scher term from YM$_2$ perturbation theory: in particular we will examine the derivation of the generating function $G(t,\epsilon,z)$. We start from the two relations (\ref{recrel})
\begin{eqnarray}\label{recrel1}
g_n(t,\epsilon)&=&\left(-\frac{g_{2d}^2\,N}{4\pi}\right)\int_\epsilon^t 
dt_n\,\Delta(t_n)\,g_{n-1}(t_n,\epsilon)\,,\\
\frac{d}{dt}g_n(t,\epsilon)&=&\left(-\frac{g_{2d}^2\,N}{4\pi}\right)\,\Delta(t)\,g_{n-1}(t,\epsilon)
\end{eqnarray}
with $g_0(t,\epsilon)=1$.
Using the identity
\begin{equation}
\Delta(t)=\frac 12 \frac{d}{dt}\log\left(\frac{t^2}{t^2+1}\right)
\end{equation}
into the recurrence relation \eqref{recrel1} and integrating by parts we obtain
\begin{equation}\begin{split}\label{intpart1}
g_n(t,\epsilon)=&\left(-\frac{g_{2d}^2\,N}{8\pi}\right)
\Biggl\{\log\left(\frac{t^2}{t^2+1}\right)g_{n-1}(t,\epsilon)-
\log\left(\frac{\epsilon^2}{\epsilon^2+1}\right)g_{n-1}(\epsilon,\epsilon)-\\
&\qquad\qquad\qquad-\!\int_\epsilon^t 
dt_n\,\log\left(\frac{t_n^2}{t_n^2+1}\right)\,\frac{d}{dt_n}g_{n-1}(t_n,\epsilon)\Biggl\}\\
=&\left(-\frac{g_{2d}^2\,N}{8\pi}\right)
\Biggl\{\log\left(\frac{t^2}{t^2+1}\right)g_{n-1}(t,\epsilon)-
\log\left(\frac{\epsilon^2}{\epsilon^2+1}\right)g_{n-1}(\epsilon,\epsilon)-\\
&\qquad\qquad-\left(-\frac{g_{2d}^2\,N}{4\pi}\right)\!\int_\epsilon^t 
dt_n\,\log\left(\frac{t_n^2}{t_n^2+1}\right)\,\Delta(t_n)\,g_{n-2}(t_n,\epsilon)\Biggl\}\,.
\end{split}\end{equation}
We rewrite the product in the last line as
\begin{equation}
\log\left(\frac{t_n^2}{t_n^2+1}\right)\,\Delta(t_n)=\frac{1}{4}\frac{d}{dt_n}\log^2 
\left(\frac{t_n^2}{t_n^2+1}\right)
\end{equation}
and integrating by parts again we obtain
\begin{align}\label{intpart2}
g_n(t,\epsilon)=&\Biggl\{\!\Biggr[\!\left(-\frac{g_{2d}^2\,N}{8\pi}\right)
\log\left(\frac{t^2}{t^2+1}\right)g_{n-1}(t,\epsilon)-\frac{1}{2}
\left(-\frac{g_{2d}^2\,N}{8\pi}\right)^2
\log^2\left(\frac{t^2}{t^2+1}\right)g_{n-2}(t,\epsilon)\nonumber\\ &- (t\rightarrow\epsilon)\Biggr]
+\frac{1}{2}\left(-\frac{g_{2d}^2\,N}{8\pi}\right)^2\int_\epsilon^t 
dt_n\,\log^2\left(\frac{t_n^2}{t_n^2+1}\right)\,\frac{d}{dt_n}g_{n-2}(t_n,\epsilon)\Biggl\}\,.
\end{align}
The procedure can be iterated $n-1$ times and, defining $\alpha=-\frac{g_{2d}^2\,N}{8\pi}$, we arrive at
\begin{equation}\label{recurrvera1}
g_n(t,\epsilon)=-\sum_{k=1}^n\frac{(-\alpha)^k}{k!}\log^k\left(\frac{t^2}{t^2+1}\right)\,
g_{n-k}(t,\epsilon)-(t\rightarrow \epsilon)\,.
\end{equation} 
We can solve the recurrence by finding the generating function:
given the sequence $g_n(t,\epsilon)$ we define
\begin{equation}
G(t,\epsilon,z)=\sum_{n=0}^\infty g_n(t,\epsilon)\,z^n\qquad\text{with}\quad z\in 
\mathbb{C}\,,
\end{equation}
then using Cauchy's formula  
\begin{equation}
g_n(t,\epsilon)=\frac{1}{2\pi i}\oint_\gamma \frac{dz}{z^{n+1}}G(t,\epsilon,z)
\end{equation}
with $\gamma$ a closed curve around the origin.

Going back to the equation \eqref{recurrvera1}, we notice that $g_{n-k}(\epsilon,\epsilon)=0$ 
for $k\neq n$. Then we get:
\begin{align}
G(t,\epsilon,z)-1=&\sum_{n=1}^\infty g_n(t,\epsilon)z^n=
-\sum_{n=1}^\infty z^n\sum_{k=1}^n\frac{(-\alpha)^k}{k!}\log^k\left(\frac{t^2}{t^2+1}\right)\,
g_{n-k}(t,\epsilon)+\nonumber\\
&\phantom{\sum_{n=1}^\infty g_n(t,\epsilon)z^n=}+\sum_{n=1}^\infty 
\frac{(-\alpha z)^n}{n!}\log^n\left(\frac{\epsilon^2}{\epsilon^2+1}\right)\nonumber\\
=&-\sum_{k=1}^\infty z^k\frac{(-\alpha)^k}{k!}\log^k\left(\frac{t^2}{t^2+1}\right)\,
\sum_{n=k}^\infty g_{n-k}(t,\epsilon)z^{n-k}+e^{-\alpha z \log 
\left(\frac{\epsilon^2}{\epsilon^2+1}\right)}-1\nonumber\\
=&-\left(e^{-\alpha z \log \left(\frac{t^2}{t^2+1}\right)}-1\right)\,
G(t,\epsilon,z)+e^{-\alpha z \log 
\left(\frac{\epsilon^2}{\epsilon^2+1}\right)}-1\,.
\end{align}\normalsize  
Finally the generating function is:
\begin{equation}\label{defg}
G(t,\epsilon,z)=\left(\frac{t^2}{t^2+1}\frac{\epsilon^2+1}{\epsilon^2}\right)^{\alpha 
z}.
\end{equation}
Following the same steps, we can find that 
the generating function associated to the sequence $(-1)^n g_n(s,\epsilon)$ is $G^{-1}(s,\epsilon,w)$.

Let us now consider the expression \eqref{Wres} and (\ref{geosum}): we have to evaluate a double-integral of the following type
\begin{equation}
A\equiv \frac{1}{(2\pi i)^2}\oint dz dw \frac{z^{-(L+1)}-w^{-(L+1)}}{w-z}\, F(z,w)
\end{equation}
with $F(z,w)$ analytic around $z,w=0$ and symmetric in $w\leftrightarrow z$
\begin{equation}
F(z,w)=\sum_{n=0}^\infty\sum_{m=0}^\infty\, a_{n,m}z^n\,w^m\,,
\end{equation}
where $a_{n,m}=a_{m,n}$.
Redefining $z\rightarrow zw$ we have
\begin{equation}
A=\frac{1}{(2\pi i)^2}\oint dz dw 
\frac{1}{w^{L+1}}\frac{z^{-(L+1)}-1}{1-z}\,\sum_{m,n=0}^\infty\,a_{n,m}z^{n}w^{n+m}\,.
\end{equation}
Performing the integral over $w$, we obtain
\begin{equation}\begin{split}
A=\frac{1}{(2\pi i)}\sum_{n=0}^L\,a_{n,L-n}\,\oint 
\frac{dz}{z^{L+1-n}}\frac{1-z^{L+1}}{1-z}=\sum_{n=0}^L\,a_{n,L-n}\,.
\end{split}\end{equation}
Notice that the function $F(w,z)$ at $w=z$ has the form
\begin{equation}
F(z,z)=\sum_{n=0}^\infty\sum_{m=0}^\infty\, a_{n,m}z^{n+m}=\sum_{k=0}^\infty b_k 
z^k
\end{equation}
whit $b_k=\sum_{n=0}^k a_{n,k-n}$. Then
\begin{equation}
A=b_L=\frac{1}{2\pi i}\oint \frac{dz}{z^{L+1}}F(z,z),
\end{equation}
{\it i.e.}
\begin{equation}\label{residui42}
\oint dz dw \frac{z^{-(L+1)}-w^{-(L+1)}}{w-z}\, F(z,w)=2\pi i \oint 
\frac{dz}{z^{L+1}}F(z,z)\,.
\end{equation}

\end{document}